\documentclass[twocolumn,superscriptaddress,showpacs]{revtex4}  
\usepackage{ae}
\usepackage[T1]{fontenc}
\usepackage[ansinew]{inputenc}
\usepackage{amsmath}
\usepackage{graphicx}
\usepackage{color}
\usepackage[colorlinks]{hyperref}
\usepackage{setspace}

\begin{document}

\date{\today}
\title{Electronic spin working mechanically}
\author{R. I. Shekhter}
\affiliation{Department of Physics, University of Gothenburg, SE-412
96 G{\" o}teborg, Sweden}
\author{L. Y. Gorelik}
\affiliation{Department of Applied Physics, Chalmers University of
Technology, SE-412 96 G{\" o}teborg, Sweden}
\author{I. V. Krive}
\affiliation{B. Verkin Institute for Low Temperature Physics and
Engineering of the National Academy of Sciences of Ukraine, 47 Lenin
Ave., Kharkov 61103, Ukraine} \affiliation{Physical Department, V.
N. Karazin National University, Kharkov 61077, Ukraine}
\author{M. N. Kiselev}
\affiliation{The Abdus Salam International Centre for Theoretical
Physics, Strada Costiera 11, 1-34151 Trieste, Italy}
\author{S. I. Kulinich}
\affiliation{B. Verkin Institute for Low Temperature Physics and
Engineering of the National Academy of Sciences of Ukraine, 47 Lenin
Ave., Kharkov 61103, Ukraine}
\author{A. V. Parafilo}
\affiliation{B. Verkin Institute for Low Temperature Physics and
Engineering of the National Academy of Sciences of Ukraine, 47 Lenin
Ave., Kharkov 61103, Ukraine}
\author{K. Kikoin}
\affiliation{School of Physics and Astronomy, Tel Aviv University, 69978 Tel Aviv, Israel}
\author{M. Jonson}
\affiliation{Department of Physics, University of Gothenburg, SE-412 96 G{\" o}teborg, Sweden}
\affiliation{SUPA, Institute of Photonics and Quantum Sciences, Heriot-Watt University, Edinburgh EH14 4AS, Scotland, UK}
\affiliation{Department of Physics, Division of Quantum Phases and
Devices, Konkuk University, Seoul 143-701, Korea}

\date{\today}
\pacs{73.23.-b,
  72.10.Fk,
  73.23.Hk,
  85.85.+j}

\begin{abstract}
A single-electron tunneling (SET) device with a nanoscale central
island that can move with respect to the bulk source- and drain
electrodes allows for a nanoelectromechanical (NEM) coupling between
the electrical current through the device and mechanical vibrations
of the island. Although an electromechanical ``shuttle" instability
and the associated phenomenon of single-electron shuttling were
predicted more than 15 years ago, both theoretical and experimental
studies of NEM-SET structures are still carried out. New
functionalities based on quantum coherence, Coulomb correlations and
coherent electron-spin dynamics are of particular current interest.
In this article we present a short review of recent activities in
this area.
\end{abstract}
\maketitle

\section{Introduction} \label{int}


Electric weak links play a crucial role in modern nanoelectronics since
they offer a natural way to inject electrons into small conducting areas.
At the same time weak links of nanometer size offer new functionality
due to the mesoscopic properties of the small conductors that form such
links. Coulomb blockade of tunneling, resonant tunneling, quantum spin
coherence, spin-dependent tunneling and weak superconductivity are
just examples of new phenomena (compared to bulk transport phenomena)
that lead to new physics in nanometer sized weak electric links. Special
interest is focused on the non-equilibrium evolution of ``hot'' electrons with
voltage-controllable excess energy. Point contact spectroscopy of
elementary excitations and nanoelectromechanical shuttle instabilities
are the brightest examples of functionalities based on properties of
accelerated electrons in point contacts. The non-equilibrium nature of an
electronic system is most prominently manifested if excitation modes, which
are spatially localized in the vicinity of a weak link, interact with the ``hot''
electrons. Then even a low level of energy transfer from the electrons does
not prevent these excitations from accumulating a significant amount of
energy, with the energized electrons acting as power supply.

Single-electron tunneling (SET) transistors are nanodevices with
particularly prominent mesoscopic features.
Here, the Coulomb blockade of single-electron tunneling at low
voltage bias and temperature \cite{coulombblockade} makes Ohm's law for the
electrical conductance invalid in the sense that the electrical
current is not necessarily proportional to the voltage drop across
the device.
Instead, the current  is due to a temporally discrete set of events where
electrons tunnel quantum-mechanically one-by-one from a source to a
drain electrode via a nanometer size island (a ``quantum dot").
This is why the properties of a single electronic quantum state are
crucial for the operation of the entire device.

Since the probability for quantum mechanical tunneling is
exponentially sensitive to the tunneling distance, it follows that
the position of the quantum dot relative to the electrodes is
crucial. On the other hand the strong Coulomb forces that accompany
the discrete nanoscale charge fluctuations, which are a necessary
consequence of a current flow through the SET device, might cause a
significant deformation of the device and move the dot, hence giving
rise to a strong electro-mechanical coupling. This unique feature
makes the so-called nanoelectromechanical SET (NEM-SET) devices, where
mechanical deformation can be achieved along with electronic
operations,
to be one of the best nanoscale realizations of
electromechanical transduction.

In this review we will discuss some of the latest achievements in
the nano-electromechanics of NEM-SET devices focusing on the new
functionality that exploits { the coherence of quantum charge and
spin subsystems in their interplay with mechanical subsystem}.
By choosing
magnets as
components of the device one may, take advantage of a
macroscopic ordering of electrons with respect to
their spin. We will discuss how the
electronic spin contribute to electromechanical and
mechano-electrical transduction in a NEM-SET device. New effects
appear also due to
{ many-body reconstruction of the electron spectrum in the metallic
leads related to exchange interaction with spin localized in the
moving shuttle. This interaction opens a new channel of Kondo
resonance tunneling between the shuttle and the leads, which
contributes to specific "Kondo- nano-mechanics". }

This review is
an update
of our earlier reviews of shuttling \cite{shuttlerev1, shuttlerev2,
shuttlerevfnt}. Other aspects of nanoelectromechanics are only
briefly discussed here. We refer readers to the well-known reviews
of Refs.~\onlinecite {blencowe, roukes1, roukes2,cleland,zant} on
nanoelectromechanical systems for additional information.

\section{Shuttling of single electrons}

A single-electron shuttle can be considered as the ultimate
miniaturization of a classical electric pendulum capable of
transferring macroscopic amounts of charge between two metal plates.
In both cases the electric force acting on a charged ``ball" that is
free to move in a potential well between two metal  electrodes kept
at different electrochemical potentials, $eV=\mu_L-\mu_R$, results
in self-oscillations of the ball. Two distinct physical phenomena,
namely the quantum mechanical tunneling mechanism for charge loading
(unloading) of the ball (in this case more properly referred to as a
grain) and the Coulomb blockade of tunneling, distinguish the
nanoelectromechanical device known as a single-electron shuttle
\cite{shuttleprl} (see also \cite{electrinst}) from its classical
textbook analog. The regime of Coulomb blockade realized at bias
voltages and temperatures $eV, T \ll E_C$ (where $E_C= e^2/2C$ is
the charging energy, $C$ is the grain's electrical capacitance)
allows one to consider single electron transport through the grain.
Electron tunneling, being extremely sensitive to the position of the
grain relative to the bulk electrodes, leads to a shuttle
instability --- the absence of any equilibrium position of an
initially neutral grain in the gap between the electrodes.

\subsection{Shuttle instability in the quantum regime of Coulomb blockade}

\begin{figure}\label{fig1}
\vspace{0.cm} {\includegraphics[width=8cm]{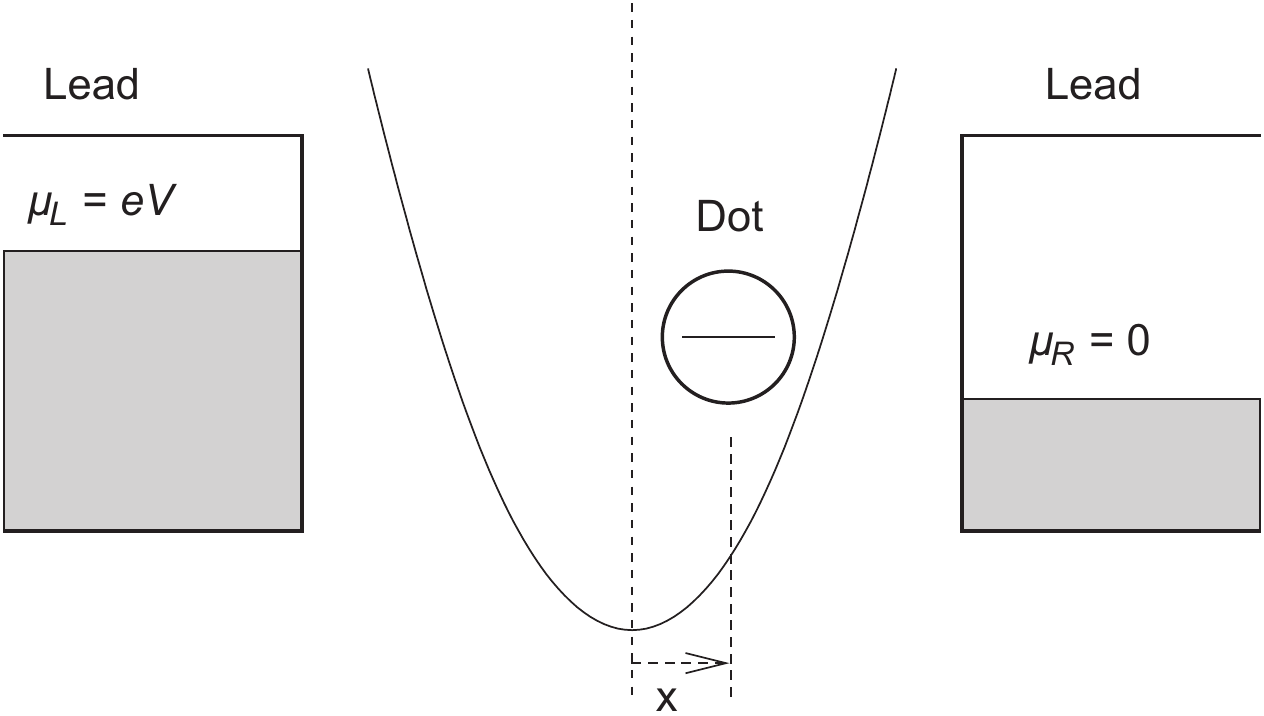}} \vspace*{0.5
cm} \caption{Model system consisting of a movable quantum dot placed
between two leads. An effective elastic force acting on the dot due
to its connections to the leads is described by a parabolic
potential. Only one single electron state is available in the dot
and the non-interacting electrons in the leads are assumed to have a
constant density of states. Reprinted with permission from
\cite{shuttlefedor}, D. Fedorets {\em et al.},  { Europhys. Lett.}
{\bf 58}, 99 (2002). $\copyright$ 2002, EDP Sciences. }
\end{figure}

At first, we consider the single-electron shuttle effect in the
simplest model \cite{shuttlefedor} where the grain is modeled as a
single-level quantum dot (QD) that is weakly coupled (via a tunnel
Hamiltonian) to the electrodes (see Fig. 1). The Hamiltonian
corresponding to this model reads

\begin{equation} \label{Model}
H_{tot}=\sum_{j=L,R}H^{(j)}_l+H_{QD}+H_v+\sum_{j=L,R}H_t^{(j)},
\end{equation}
where the Hamiltonian
\begin{equation} \label{lead}
H_l^{(j)}=\sum_k(\varepsilon_{kj}-\mu_j)a^{\dag}_{kj}a_{kj}
\end{equation}
describes noninteracting electrons in the left ($j=L$) and right
($j=R$) leads, which are kept at different chemical potential
$\mu_j$ and have a constant density states $\nu_{j}$;
$a^{\dag}_{kj}(a_{kj})$ creates (annihilates) an electron with
momentum $k$ in  lead $j$. The quantum dot is described by two
parts. It is single electron level Hamiltonian and Hamiltonian of
harmonic potential in which QD vibrates
\begin{equation} \label{singlelevel}
H_{QD}=\varepsilon_0c^{\dag}c-dxc^{\dag}c,
\end{equation}
\begin{equation} \label{vibrating}
H_{v}=\frac{1}{2}(x^2+p^2),
\end{equation}
where $c^{\dag}(c)$ is the creation (annihilation) operator for an
electron at the dot, $\varepsilon_0$ is the energy of the resonant
level, $x$ is the dimensionless coordinate operator (normalized by
the amplitude $x_0$ of zero-point fluctuations,
$x_0=\sqrt{\hbar/M\omega_0}$, $M$ is the mass of QD), $p$ is the
corresponding momentum operator ($[x,p]=i$), $\omega_0$ is the
frequency of vibrons, $d=eE/(M\omega_0^2 x_0)$ is characteristic
electromechanical interaction constant. For convenience we use
dimensionless variables. The physical meaning of the second term in
Eq.~(\ref{singlelevel}) for usual shuttle systems is the interaction
energy due to the coupling of the electron charge density on the dot
with the electric
 field ($E$)
in the gap  between electrodes. Here, for convenience, all energies
measure in units $\hbar\omega_0$, time in units of $\omega_0^{-1}$.
Note, that in general the mechanism of electromechanical interaction
could have different nature (electrostatic interaction charge on the
dot with gate electrode, interaction in magnetic field due the
Lorentz force, due exchange force between electrons with spin and
spin polarized leads, see next Sections).

The tunneling Hamiltonian $H_t^{(j)}$ in Eq.~(\ref{Model})
has the form:
\begin{equation} \label{tunnel}
H_t^{(j)}=\sum_kt_{0j}\exp(jx/\lambda)a_{kj}^{\dag}c+h.c.
\end{equation}
Here $j=\pm$ for $L/R$ electrodes, $t_{0j}$ is the bare tunneling
amplitude, which corresponds to a weak dot-electrode coupling,
$\lambda$ is the characteristic tunneling length. The explicit
coordinate dependence in the tunneling Hamiltonian indicates
sensitivity of tunnel matrix elements to a shift of the quantum dot
center-of-mass coordinate with respect to its equilibrium
($x_{cm}=0$) position. The $x$-dependence in Eq. (\ref{tunnel})
represents also additional interaction with vibronic degree of
freedom.

Even in such a simple formulation the single-electron shuttle problem is
quite complex.
In this section we review some main results of electron shuttling
(without involving the spin degree of freedom) and present the basic
idea of solution's method based on the equation of motion for the
matrix density. The advantage of this method is that it is possible
to explicitly consider the quantum dot dynamics in quantum regime
and take into account  the coherent dynamics of spin electron's
states in a magnetic field, see the next section.

The time evolution of the system is obtained from the Liouville-von
Neumann equation for the total density matrix
\begin{equation} \label{LvN eq}
i\hbar\partial_t\hat{\sigma}(t)=[H,\hat{\sigma}(t)].
\end{equation}
In order to consider the dynamics of the electronic state in the dot and the  vibronic
degrees of freedom we reduce the total density operator by tracing
over all electronic states in the leads,
$\rho(t)=\mathrm{Tr}_{leads}\{\sigma(t)\}$. We assume that electrons
in the leads are in equilibrium and that they are not affected by the
coupling to the dot. So, we factorize the density matrix,
$\sigma(t)\approx\rho(t)\otimes\sigma_{leads}$ (this approximation
is always valid for $\Gamma_{j}=2\pi\nu_{j}|t_{0j}|^2\exp[\mp
x/\lambda]\ll1$). After shifting the $x$-axis by $d/2$ we get the
system of equation of motion for the diagonal elements of density
matrix $\rho_{0}=\langle0|\rho|0\rangle$ and
$\rho_{1}=\langle1|\rho|1\rangle$, where
$|1\rangle=d^{\dag}|0\rangle$, as
\begin{eqnarray} \label{eqsys1}
&&\partial_t\rho_{0}=-i\left[
(H_v+\frac{d}{2}x,\rho_{0}\right]-\frac{1}{2}\{\overline{\Gamma}_L(x),\rho_{0}\}
\nonumber
\\
&&+\sqrt{\overline{\Gamma}_R(x)}\rho_{1}\sqrt{\overline{\Gamma}_R(x)},
\end{eqnarray}
\begin{eqnarray} \label{eqsys2}
&&\partial_t\rho_{1}=-i\left[
(H_v-\frac{d}{2}x,\rho_{1}\right]-\frac{1}{2}\{\overline{\Gamma}_R(x),\rho_{1}\}
\nonumber
\\
&&+\sqrt{\overline{\Gamma}_L(x)}\rho_{0}\sqrt{\overline{\Gamma}_L(x)},
\end{eqnarray}
where $\overline{\Gamma}_j(x)=\Gamma_j (x+d/2)]$. The off-diagonal
density matrix elements are decoupled from the equation of motion
of the diagonal elements. It is easy to take into account
dissipation of the system. The corresponding dissipation term is $L
_{\gamma}\rho=-(i\gamma/2)[x,\{ p,\rho \}]-(\gamma/2)[x,[x,p]]$
($\gamma$ is the dissipation rate).

Now we find the condition under which the vibrational ground state of
the oscillator becomes unstable. For this we consider the time
evolution of the expectation value of the coordinate,
$\bar{x}(t)=\textbf{Tr}\{x\rho_{+}\}$, and the momentum operators,
$\bar{p}(t)= \textbf{Tr}\{p\rho_{+}\}$, of the island (here
$\rho_{+}\equiv\rho_0+\rho_1$). To first order in $\lambda^{-1}$,
for symmetric tunneling couplings
$\widetilde{\Gamma}_L(0)=\widetilde{\Gamma}_R(0)=\Gamma/2$ and in
the high bias voltage limit ($\mu_L-\mu_R=eV\rightarrow\infty$) the
equations of motion for the first vibrational moments become closed, so that
\cite{quanshuttle}
\begin{eqnarray} \label{sys}
&\dot{\bar{x}}=\bar{p}, &
\dot{\bar{p}}=-\gamma\bar{p}-\bar{x}-\frac{d}{2}n_{-},
\\
&& \dot{n}_{-}=-\Gamma
n_{-}+\frac{2\Gamma}{\lambda}\bar{x},\nonumber
\end{eqnarray}
where $n_{-}=1-2 \bf{Tr}\rho_{1}$. The solution of Eq. (\ref{sys})
for the quantum dot displacement is $\bar{x}(t)\approx A e^{rt}\cos(t)$,
where $r=1/2(\gamma_{thr}-\gamma)$ is the rate of increment of the shuttle
instability. If the dissipation rate $ \gamma$ is below the threshold
value $\gamma_{thr}=\Gamma d/[\lambda(\Gamma^2)+1]$, then the
expectation value of the dot coordinate grows exponentially in time
and the vibrational ground state is unstable. It was shown
\cite{quanshuttle} that this exponential increase of the displacement
drives the system into the nonlinear regime of the vibration
dynamics, where the system reaches a stable steady state of
developed shuttle motion.

In order to analyze this stable state (i.e. the solution of the
system Eq. (\ref{eqsys1},\ref {eqsys2}) it is convenient to use the
Wigner function representation \cite{novotny}, \cite{quanshuttle}.
The Wigner distribution function for the density operator $\rho_{+}$
is defined as
\begin{equation} \label{wigner}
W_{+}(x,p)\equiv\frac{1}{2\pi}\int_{-\infty}^{+\infty}d\xi
e^{-ip\xi}\left\langle x+\xi/2|\rho_{+}|x-\xi/2\right\rangle.
\end{equation}
The dynamics of the oscillating QD is characterized by its
trajectory (distribution) in the phase space ($x, p$) for
$p^2/2+x^2/2=const$. Now we proceed to polar coordinates
$(A,\varphi)$, where $x=A\sin\varphi$ and $p=A\cos\varphi$.
An equation for $W_{+}(A, \varphi)$ is derived from Eqs. (\ref{eqsys1})
and (\ref{eqsys2}) after straightforward calculations (for details see
\cite{quanshuttle}). To leading order in the small parameters
$d/\lambda$, $\lambda^{-2}$, and $\gamma$ this equation takes the form
of a stationary Fokker-Planck equation for the zeroth Fourier
component of the Wigner function $\overline{W}_{+}(A)$
\begin{equation} \label{FPeq}
\frac{\partial}{\partial
A}\left(\overline{D}_0(A)\frac{\partial}{\partial
A}-\overline{D}_1(A)\right)\overline{W}_{+}(A)=0,
\end{equation}
where $\overline{D}_1=A^2 D_1(A), \overline{D}_0=A D_0(A)$ are
drift- and diffusion coefficients  (analytical expression of this
coefficients will be presented in section for the magnetic shuttle).
The normalized solution of Eq. (\ref{FPeq}) has the form of a Boltzman
distribution,
\begin{equation} \label{Sol}
\overline{W}_{+}=Z^{-1}\exp\left(\int_0^AdA \frac{\overline{
D}_1(A)}{\overline{D}_0(A)}\right)
\end{equation}
The stationary solution of the oscillating dot is localized in the
phase space around points where $\overline{W}_{+}$ is maximal. From Eq.
(\ref{Sol}) one can see that the maximum of the Wigner function is
determined by zeros of the drift coefficient $\overline{D}_1(A_m)=0$
($\overline{D}'_1(A_m)<0$). In the vicinity of this point,
$\overline{W}_{+}$ can be approximated by a Gaussian distribution
function. For the spinless shuttle problem it can be shown that
$\overline{W}_{+}$ always has an extremum at $A=0$: maximum for
$\gamma>\gamma_{thr}$ and minimum for $\gamma<\gamma_{thr}$. So
the vibrational ground state is unstable when the dissipation is below
threshold value as has been shown by solving the equation system
(\ref{sys}). The function $\overline{W}_{+}$ has also a maximum for
the non-zero amplitude $A_C$, which corresponds to the stable limit cycle
amplitude of shuttle oscillations (for more details see
\cite{quanshuttle}).

 One can
distinguish two regimes of "quantum" (for
$d/\lambda\ll\lambda^{-4}$) and "quasiclassical"
($d/\lambda\gg\lambda^{-4}$) shuttle motion.  In the quasiclassical
regime Gaussian distribution is narrow and in quantum regime the
width of distribution ``bell'' is of the order of $\lambda\gg 1$, i.e. the
Wigner function is smeared around classical phase trajectory. It is
interesting to note that a region of parameters exists where both
vibrational and shuttling regimes are present (a region
where the Wigner function has two maxima).

\section{Electro - and Spintro - Mechanics of Magnetic shuttle devices} \label{c}

In this Section we will explore new functionalities that
emerge when nanomechanical devices are partly or completely made of
magnetic materials. The possibility of magnetic ordering
brings new degrees of freedom into play in addition to the
electronic and mechanical ones considered so far, opening up an
exciting perspective towards utilising magneto-electro-mechanical
transduction for a large variety of applications. Device dimensions
in the nanometer range mean that a number of mesoscopic phenomena in
the electronic, magnetic and mechanical subsystems can be used for
quantum coherent manipulations. In comparison with the
electromechanics of the nanodevices considered above the prominent
role of the electronic spin in addition to the electric charge
should be taken into account.

The ability to manipulate and control spins via electrical \cite{9,
10, 11} magnetic \cite{12} and optical \cite{kadigrobov, 13} means
has generated numerous applications in metrology \cite{14} in recent
years. A promising alternative method for spin manipulation employs
a mechanical resonator coupled to the magnetic dipole moment of the
spin(s), a method which could enable scalable quantum information
architectures \cite{15} and sensitive nanoscale magnetometry
\cite{16, 17, 18}. Magnetic resonance force microscopy (MRFM) was
suggested as a means to improve spin detection to the level of a
single spin and thus enable three dimensional imaging of
macromolecules with atomic resolution. In this technique a single
spin, driven by a resonant microwave magnetic field interacts with a
ferromagnetic particle. If the ferromagnetic particle is attached to
a cantilever tip, the spin changes the cantilever vibration
parameters \cite{21}. The possibility to detect \cite{21} and
monitor the coherent dynamics of a single spin mechanically
\cite{22} has been demonstrated experimentally. Several theoretical
suggestions concerning the possibility to test single-spin dynamics
through an electronic transport measurement were made recently
\cite{24, 25, 26, 27}. Complementary studies of the mechanics of a
resonator coupled to spin degrees of freedom by detecting the spin
dynamics and relaxation were suggested in
\cite{24,25,26,27,28,29,30,31} and carried out in \cite{32}.
Electronic spin-orbit interaction in suspended nanowires was shown
to be an efficient tool for detection and cooling of bending-mode
nanovibrations as well as for manipulation of spin qubit and
mechanical quantum vibrations \cite{33, 34, 35}.

An obvious modification of the nano-electro-mechanics of magnetic
shuttle devices originates from the spin-splitting of electronic
energy levels, which results in the known phenomenon of
spin-dependent tunneling. Spin-controlled nano-electro-mechanics
which originates from spin-controlled transport of electric charge
in magnetic NEM systems is represented by number of new
magneto-electro-mechanical phenomena.

Qualitatively new opportunities appear when magnetic nanomechanical
devices are used. They have to do with the effect of the
short-ranged magnetic exchange interaction between the spin of
electrons and magnetic parts of the device. In this case the spin of
the electron rather than its electrical charge can be the main
source of the mechanical force acting on movable parts of the
device. This leads to new physics compared with the usual
electromechanics of non-magnetic devices, for which we use the term
spintro-mechanics. In particular it becomes possible for a movable
central island to shuttle magnetization between two magnetic leads
even without any charge transport between the leads. The result of
such a mechanical transportation of magnetization is a magnetic
coupling between nanomagnets with a strength and sign that are
mechanically tunable.

In this Section we will review some early results that involve the
phenomena mentioned above. These only amount to a first step in the
exploration of new opportunities caused by the interrelation between
charge, spin and mechanics on a nanometer length scale.

\subsection{ Spin-controlled shuttling of electric charge} \label{w}

By manipulating the interaction between the spin of electrons and
external magnetic fields and/or the internal interaction in magnetic
materials, spin-controlled nanoelectromechanics may be achieved.

A new functional principle --- spin-dependent shuttling of electrons
--- for low magnetic field sensing purposes was proposed by Gorelik
{\em et al.} in Ref.~\onlinecite{r88}. This principle may lead to a giant
magnetoresistance effect in  external magnetic fields as low as
1-10~Oe in a magnetic shuttle device if magnets with highly
spin-polarized electrons (half metals \cite{r83, r84, r85, r86,
r87}) are used as leads in a magnetic shuttle device. The key idea
is to use the external magnetic field to manipulate the spin of
shuttled electrons rather than the magnetization of the leads. Since
the electron spends a relatively long time on the shuttle, where it
is decoupled from the magnetic environment, even a weak magnetic can
rotate its spin by a significant angle. Such a rotation allows the
spin of an electron that has been loaded onto the shuttle from a
spin-polarized source electrode to be reoriented in order to allow
the electron finally to tunnel from the shuttle to the (differently)
spin-polarized drain lead. In this way the shuttle serves as a very
sensitive ``magnetoresistor" device. The model employed in
Ref.~\onlinecite{r88} assumes that the source and drain are fully polarized
in opposite directions. A mechanically movable quantum dot
(described by a time-dependent displacement $x(t)$), where a single
energy level is available for electrons, performs driven harmonic
oscillations between the leads. The external magnetic field, $H$, is
perpendicular to the orientations of the magnetization in both leads
and to the direction of the mechanical motion.

The spin-dependent part of the Hamiltonian is specified as
\begin{equation} \label{Spin trans ham}
H_{\rm
magn}(t)=J(t)(a^{\dag}_{\uparrow}a_{\uparrow}-a^{\dag}_{\downarrow}a_{\downarrow})
-\frac{g\mu
H}{2}(a^{\dag}_{\uparrow}a_{\downarrow}+a^{\dag}_{\downarrow}a_{\uparrow}),
\end{equation}
where $J(t)=J_R(t)-J_L(t)$, $J_{L(R)}(t)$ are the { molecular fields
induced by exchange interactions} between the on-grain electron and
the left(right) lead, $g$ is the gyromagnetic ratio and $\mu$ is the
Bohr magneton. The proper Liouville-von Neumann equation for the
density matrix is analyzed and an average electrical current is
calculated for the case of large bias voltage.

In the limit of weak exchange field, $J_{max}\ll \mu H$ one
may neglect the influence of the magnetic leads on the on-dot
electron spin dynamics. The resulting current is
\begin{equation} \label{Mech assi curr}
I=\frac{e \omega_0}{\pi}\frac{\sin^2(\vartheta/2)\tanh(w/4)}{\sin^2
(\vartheta/2)+\tanh^2(w/4)}
\end{equation}
where $w$ is the total tunneling probability during the contact time
$t_0$, while $\vartheta\sim\pi g\mu H/\hbar\omega_0$ is the rotation
angle of the spin during the ``free-motion" time.

The theory \cite{r88} predicts oscillations in the magnetoresistance
of the magnetic shuttle device with a period $\Delta H_p$, which is
determined from the equation $\hbar\omega_0=g \mu (1+w)\Delta H_p$.
The physical meaning of this relation is simple: every time when
$\omega_0/\Omega=n+1/2$ ($\Omega=g \mu H/\hbar$ is the spin
precession frequency in a magnetic field) the shuttled electron is
able to flip fully its spin to remove the ``spin-blockade" of
tunneling between spin polarized leads having their magnetization in
opposite directions. This effect can be used for measuring the
mechanical frequency thus providing dc spectroscopy of
nanomechanical vibrations.

Spin-dependent shuttling of electrons as discussed above is a
property of non-interacting electrons, in the sense that tunneling
of different electrons into (and out of) the dot are independent
events. The Coulomb blockade phenomenon adds a strong correlation of
tunneling events, preventing fluctuations in the occupation of
electronic states on the dot. This effect crucially changes the
physics of spin-dependent tunneling in a magnetic NEM device. One of
the
remarkable consequences is 
the Coulomb promotion of spin-dependent tunneling predicted in
Ref.~\onlinecite{1}. In this work a strong voltage dependence of the
spin-flip relaxation rate on a quantum dot was demonstrated. Such
relaxation, being very sensitive to the occupation of spin-up and
spin-down states on the dot, can be controlled by the Coulomb
blockade phenomenon. It was shown in Ref.~\onlinecite{1} that by lifting
the Coulomb blockade one stimulates occupation of both spin-up and
spin-down states thus suppressing spin-flip relaxation on the dot.
In magnetic devices with highly spin-polarized electrons electronic
spin-flip can be the only mechanism providing charge transport
between oppositely magnetized leads. In this case the onset of
Coulomb blockade, by increasing the spin-flip relaxation rate,
stimulates charge transport through a magnetic SET device (Coulomb
promotion of spin-dependent tunneling). Spin-flip relaxation
also modifies qualitatively the noise characteristics of
spin-dependent single-electron transport. In Refs.~\onlinecite{2, 3} it was
shown that the low-frequency shot noise in such structures diverges
as the spin relaxation rate goes to zero. This effect provides an
efficient tool for spectroscopy of extremely slow spin-flip
relaxation in quantum dots. Mechanical transportation of a
spin-polarized dot in a magnetic shuttle device provides new
opportunities for studying spin-flip relaxation in quantum dots. The
reason can be traced to a spin-blockade of the mechanically aided
shuttle current that occurs in devices with highly polarized and
colinearly magnetized leads. As was shown in Ref.~\onlinecite{4} the
above effect results in giant peaks in the shot-noise spectral
function, wherein the peak heights are only limited by the rates of
electronic spin flips. This enables a nanomechanical spectroscopy of
rare spin-flip events, allowing spin-flip relaxation times as long
as $10~\mu$s to be detected.

The spin-dependence of electronic tunneling in magnetic NEM devices
permits an external magnetic field to be used for manipulating not
only electric transport but also the mechanical performance of the
device. This was demonstrated in Refs.~\onlinecite{s90, 5}. A theory of
the quantum coherent dynamics of mechanical vibrations, electron
charge and spin was formulated and the possibility to trigger a
shuttle instability by a relatively weak magnetic field was
demonstrated. It was shown that the strength of the magnetic field
required to control nanomechanical vibrations decreases with an
increasing tunnel resistance of the device and can be as low as 10
Oe for giga-ohm tunnel structures.

A new type of nanoelectromechanical self excitation caused entirely
by the spin splitting of electronic energy levels in an external
magnetic field was predicted in Ref.~\onlinecite{6} for a suspended
nanowire, where mechanical motion in a magnetic field induces an
electromotive coupling between electronic and vibrational degrees of
freedom. It was shown that a strong correlation between the
occupancy of the spin-split electronic energy levels in the nanowire
and the velocity of flexural nanowire vibrations provides energy
supply from the source of DC current, flowing through the wire, to
the mechanical vibrations thus making possible stable,
self-supporting bending vibrations. Estimations made in
Ref.~\onlinecite{6} show that in a realistic case the vibration amplitude
of a suspended carbon nanotube (CNT) of the order of 10~nm can be
achieved if magnetic field of 10~T is applied.

\subsection{Spintro-mechanics of magnetic shuttle devices}

New phenomena, qualitatively different from the electromechanics of
nonmagnetic shuttle systems, may appear in magnetic shuttle devices
in a situation when short-range magnetic exchange forces become
comparable in strength to the long-range electrostatic forces
between the charged elements of the device \cite{6}.  There is
convincing evidence that the exchange field can be several tesla at
a distance of a few nanometers from the surface of a ferromagnet
\cite{radic4, radic5, radic6, radic9}. Because of the exponential
decay of the field this means that the force experienced by a
single-electron spin in the vicinity of magnetic electrodes can be
very large. These spin-dependent exchange forces can lead to various
``spintro-mechanical" phenomena.

Mechanical effects produced by a long-range electrostatic force and
short-ranged exchange forces on a movable quantum dot are
illustrated in Fig.~2.
\begin{figure}
\vspace{0.cm} \centerline {\includegraphics[width=8cm]{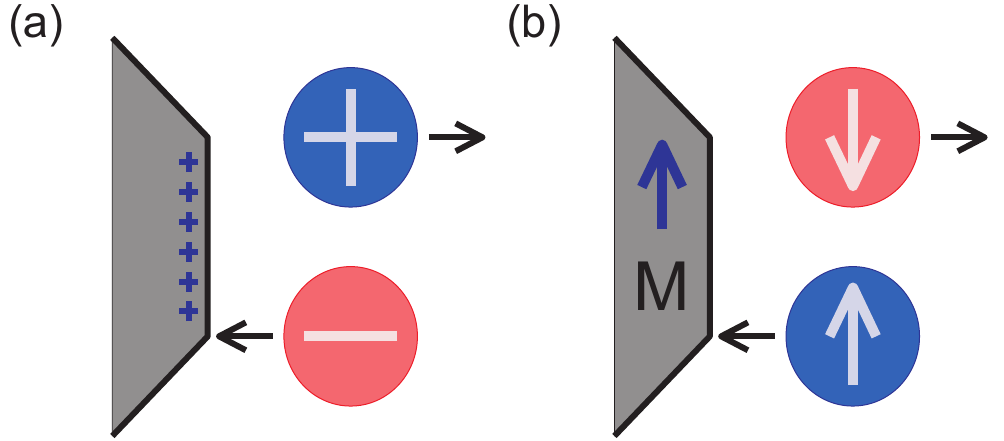}}
\vspace*{-0. cm} \caption{A movable quantum dot in a magnetic
shuttle device can be displaced in response to two types of force:
(a) a long-range electrostatic force causing an electromechanical
response if the dot has a net charge, and (b) a short-range magnetic
exchange force leading to ``spintromechanical" response if the dot
has a net magnetization (spin). The direction of the force and
displacements depends on the relative signs of the charge and
magnetization, respectively. Reprinted with permission from
\cite{7}, R. I. Shekhter {\em et al.},  {Phys. Rev. B} {\bf 86},
 100404 (2012). $\copyright$ 2012, American Physical Society.}
\end{figure}
The electrostatic force acting on the dot, placed in the vicinity of
a charged electrode (Fig.~2(a)), is determined by the electric
charge accumulated on the dot. In contrast, the exchange force
induced by a neighboring magnet depends on the net spin accumulated
on the dot. While the electrostatic force changes its direction if
the electric charge on the dot changes its sign, the spin-dependent
exchange force is insensitive to the electric charge but it changes
direction if the electronic spin projection changes its sign. A very
important difference between the two forces is that the
electrostatic force changes only as a result of injection of
additional electrons into (out of) the dot while the spintronic
force can be changed due to the electron spin dynamics even for a
fixed number of electrons on the dot (as is the case if the dot and
the leads are insulators). In this case interesting opportunities
arise from the possibility of transducing the dynamical variations
of electronic spin (induced, e.g., by magnetic or microwave fields)
to mechanical displacements in the NEM device. In
Ref.~\onlinecite{7} a particular spintromechanical effect was
discussed -- a giant spin-filtering of the electron current (flowing
through the device) induced by the formation of what we shall call a
``spin-polaronic state".

The Hamiltonian that describes the magnetic nanomechanical SET
device in Ref.~\onlinecite{7} has the standard form (its spin-dependent
part depends now on the mechanical displacement of the dot). Hence
$H=H_{lead}+H_{tunnel}+H_{dot}$, where
$H_{leads}=\sum_{k,\sigma,s}\varepsilon_{ks\sigma}a^{\dag}_{ks\sigma}a_{ks\sigma}$
describes electrons (labeled by wave vector $k$ and spin
$\sigma=\uparrow,\downarrow$) in the two leads ($s=L, R$). Electron
tunneling between the leads and the dot is modeled as
\begin{equation} \label{Spintro Tunn}
H_{tunnel}=\sum_{k,\sigma, s}T_s(x)a^{\dag}_{k
s\sigma}c_{\sigma}+H.c.
\end{equation}
where the matrix elements $T_s(x)=T^{(0)}_s\exp(\mp x/\lambda)$
($\lambda$ is the characteristic tunneling length) depend on the dot
position $x$. The Hamiltonian of the movable single-level dot is
\begin{equation} \label{Spintro dot}
H_{dot}=\hbar\omega_0b^{\dag}b+\sum_{\sigma}n_{\sigma}[\varepsilon_0-{\rm sgn}(\sigma)J(x)]+U_Cn_{\uparrow}n_{\downarrow},
\end{equation}
where ${\rm sgn}(\uparrow,\downarrow)=\pm1$, $U_C$ is the Coulomb
energy associated with double occupancy of the dot and the
eigenvalues of the electron number operators $n_{\sigma}$ is $0$ or
$1$. The position dependent magnitude $J(x)$ of the spin dependent
shift of the electronic energy level on the dot is due to the
exchange interaction with the magnetic leads. Here we expand $J(x)$
to linear order in $x$ so that $J(x)=J^{(0)}+j x$ and without loss
of generality assume that $J^{(0)}=0$.

\begin{figure}
\vspace{0.cm} \centerline {\includegraphics[width=6cm]{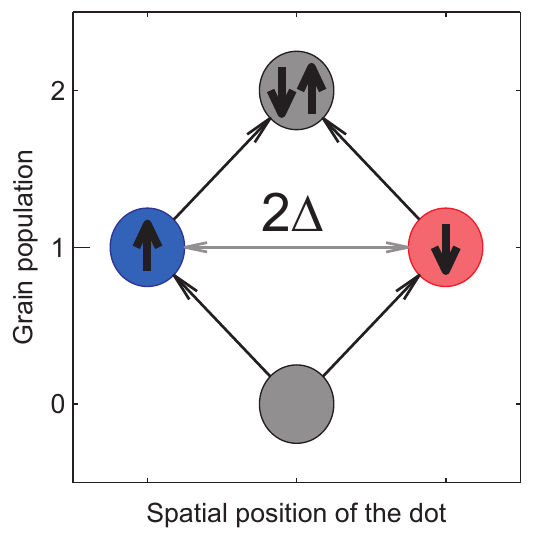}}
\vspace*{-0. cm} \caption{Diagram showing how the equilibrium
position of the movable dot depends on its net charge and spin. The
difference in spatial displacements discriminates transport through
a singly occupied dot with respect to the electron spin. Reprinted
with permission from \cite{7}, R. I. Shekhter {\em et al.}, {Phys.
Rev. B} {\bf 86},
 100404 (2012). $\copyright$ 2012, American Physical Society.}
\end{figure}

The modification of the exchange force, caused by changing the spin
accumulated on the dot, shifts the equilibrium position of the dot
with respect to the magnetic leads of the device. Since the electron
tunneling matrix element is exponentially sensitive to the position
of the dot with respect to the source and drain electrodes one
expects a strong spin-dependent renormalization of the tunneling
probability, which exponentially discriminates between the
contributions to the total electrical current from electrons with
different spins. This spatial separation of dots with opposite spins
is illustrated in Fig.~3. While changing the population of spin-up
and spin-down levels on the dot (by changing e.g. the bias voltage
applied to the device) one shifts the spatial position $x$ of the
dot with respect to the source/drain leads. It is important that the
Coulomb blockade phenomenon prevents simultaneous population of both
spin states.
If the Coulomb blockade is lifted the two spin states become equally
populated with a zero net spin on the dot, $\textbf{S}=0$. This
removes the spin-polaronic deformation and the dot is situated at
the same place as a non-populated one. In calculations a strong
modification of the vibrational states of the dot, which has to do
with a shift of its equilibrium position, should be taken into
account. This results in a so-called Franck-Condon blockade of
electronic tunneling \cite{116, ratner}. The spintro-mechanical
stimulation of a spin-polarized current and the spin-polaronic
Franck-Condon blockade of electronic tunneling are in competition
and their interplay determines a non-monotonic voltage dependence of
the giant spin-filtering effect.

To understand the above effects in more detail consider the
analytical results of Ref.~\onlinecite{7}. A solution of the problem can
be obtained by the standard sequential tunneling approximation and
by solving a Liouville equation for the density matrix for both the
electronic and vibronic subsystems. The spin-up and spin-down
currents can be expressed in terms of transition rates (energy
broadening of the level) and the occupation probabilities for
the dot electronic states.
For simplicity we consider the case of a strongly asymmetric
tunneling device. At low bias voltage and low temperature the
partial spin current is
\begin{equation} \label{Spint curr1}
I_{\sigma}\sim \frac{e
\Gamma_L}{\hbar}\exp\left(\frac{1}{2}\left[\frac{x_0^2}{\lambda^2}-\left(\frac{x_0}
{\hbar\omega_0}\right)^2\right]-{\rm sgn}(\sigma)\beta\right),
\end{equation}
where $\beta=x_0^2/\hbar\omega_0\lambda$. In the high bias voltage
(or temperature) regime, $max\{eV,T\gg E_p\}$, where the polaronic
blockade is lifted (but double occupancy of the dot is still
prevented by the Coulomb blockade), the current expression takes the
form
\begin{equation} \label{Spintr curr2}
I_{\sigma}\sim \frac{e
\Gamma_L}{\hbar}\exp\left(\left[2n_B+1\right]\frac{x_0^2}{\lambda^2}-2\,
{\rm sgn}(\sigma)\beta\right),
\end{equation}
where $n_B$ is Bose-Einstein distribution function. The scale of the
polaronic spin-filtering of the device is determined by the ratio
$\beta$ of the polaronic shift of the equilibrium spatial position
of a spin-polarized dot and the electronic tunneling length. For
typical values of the exchange interaction and mechanical properties
of suspended carbon nanotubes this parameter is about 1-10. As was
shown this is enough for the spin filtering of the electrical
current through the device to be nearly 100 \% efficient. The
temperature and voltage dependence of the spin-filtering effect is
presented in Fig.~4. The spin filtering effect and the Franck-Condon
blockade
\begin{figure}
\vspace{0.cm} \centerline {\includegraphics[width=8cm]{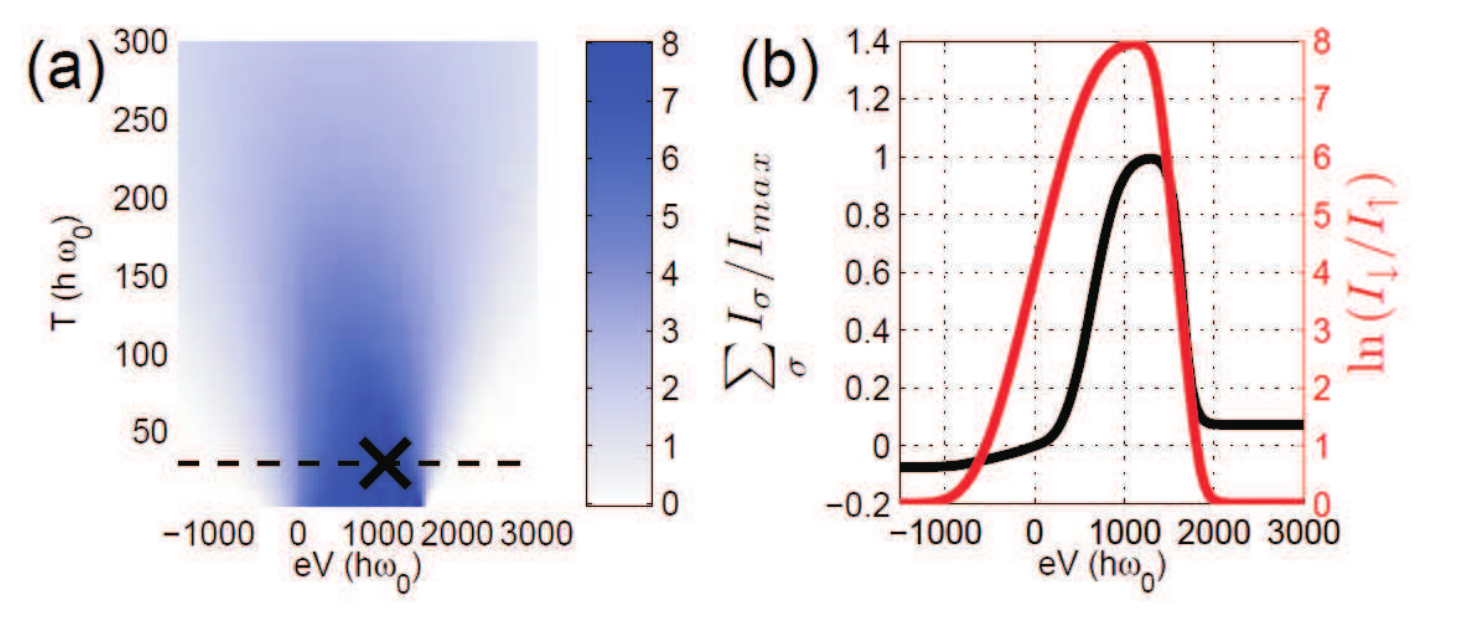}}
\vspace*{-0. cm} \caption{ Spin polarization of the current through
the model NEM-SET device under discussion. Reprinted with permission
from \cite{7}, R. I. Shekhter {\em et al.}, {Phys. Rev. B} {\bf 86},
 100404 (2012). $\copyright$ 2012, American Physical Society.}
\end{figure}
both occur at low voltages and temperatures (on the scale of the
polaronic energy; see Fig.~4 (a)).  An increase of the voltage
applied to the device lifts the Franck-Condon blockade, which
results in an exponential increase of both the current and the
spin-filtering efficiency of the device. This increase is blocked
abruptly at voltages for which the Coulomb blockade is lifted. At
this point a double occupation of the dot results in spin
cancellation and removal of the spin-polaronic segregation. This
leads to an exponential drop of both the total current and the spin
polarization of the tunnel current (Fig.~4~(b)). As one can see in
Fig.~4 prominent spin filtering can be achieved for realistic device
parameters. The temperature of operation of the spin-filtering
device is restricted from above by the Coulomb blockade energy. One
may, however, consider using functionalized nanotubes
\cite{pulkin20} or graphene ribbons \cite{pulkin21} with one or more
nanometer-sized metal or semiconductor nanocrystal attached. This
may provide a Coulomb blockade energy up to a few hundred kelvin,
making spin filtering a high temperature effect \cite{7}.

\subsection{Spintronics of shuttles}

In this subsection we discuss the possibility to manipulate the spin of
tunneling electrons by an external magnetic field and how it can affect
electron transport through a nanoelectromechanical device. In the
simplest model, we assume that the left and right electrodes are
fully spin polarized. The movable single level quantum dot (in the
absence of a magnetic field) can vibrate in the gap between two leads.
A bias voltage is applied but electron transport through the system
is blocked since the source and drain leads are fully spin polarized
in opposite direction.
An external magnetic field applied pendicular to the direction of the
magnetization in the electrode leads to precession of the electron spin of the
quantum dot and as a consequence the electron transport is unblocked. The
Hamiltonian of the system has the form \cite{5} of Eq. (\ref{Model})
with $H_{leads}=\Sigma_{jk}\varepsilon_{j}(k) c^{\dag}_{jk}c_{jk}$
($j=L, R \rightarrow j=(\uparrow,\downarrow)$) and
\begin{equation} \label{spint}
H_{QD}=(\varepsilon_0-dx)\sum_{\sigma}
c^{\dag}_{\sigma}c_{\sigma}-\frac{h}{2}(c^{\dag}_{\uparrow}c_{\downarrow}+
c^{\dag}_{\downarrow}c_{\uparrow})+Uc^{\dag}_{\uparrow}c_{\uparrow}c^{\dag}_{\downarrow}c_{\downarrow},
\end{equation}
where $h=g\mu_B H/\hbar\omega_0$ is the dimensionless magnetic field.
To analyze this system we use the method described in Section 2.
A quantum master equation for the reduced density matrix operator
$\rho_{0}\equiv \langle0\mid\rho\mid0\rangle$, $\rho_{\uparrow}
\equiv \langle\uparrow\mid\rho\mid\uparrow\rangle$,
$\rho_{\downarrow}\equiv
\langle\downarrow\mid\rho\mid\downarrow\rangle$, and
$\rho_{\uparrow\downarrow}\equiv
\langle\uparrow\mid\rho\mid\downarrow\rangle$ is obtained in analogy with
the spinless case
\begin{eqnarray}
&&\frac{\partial \rho_0}{\partial t}=-i\left[{ H}_{v}+x d,
\rho_0\right]\nonumber\label{m01}\\
&&\hspace{1.1cm}-\left\{\overline{\Gamma}_L(x),\rho_0\right\}/2
+\sqrt{\overline{\Gamma}_R(x)}\rho_\downarrow\sqrt{\overline{\Gamma}_R(x)},\\
&&\frac{\partial \rho_\downarrow}{\partial t}=-i\left[{
H}_{v},\rho_\downarrow\right]\nonumber\label{m02}\\&&\hspace{1.1cm}+i\frac{h}{2}\left(\rho_{\uparrow\downarrow}
-\rho_{\uparrow\downarrow}\right)-\frac{1}{2}\left\{{\overline{\Gamma}_{+}(x)},\rho_{\downarrow}\right\}
,\\&& \frac{\partial \rho_\uparrow}{\partial t}=-i\left[{
H}_{v},\rho_\uparrow\right]-i\frac{h}{2}\left(\rho_{\uparrow\downarrow}
-\rho_{\uparrow\downarrow}\right)
\nonumber\label{m03}\\&&\hspace{1.1cm}
+\sqrt{\overline{\Gamma}_L(x)}\rho_0\sqrt{\overline{\Gamma}_L(x)}+\sqrt{\overline{\Gamma}_R(x)}\rho_2\sqrt{\overline{\Gamma}_R(x)},
\\&& \frac{\partial \rho_{\uparrow\downarrow}}{\partial t}=
-i\left[{ H}_{v},
\rho_{\uparrow\downarrow}\right]+i\frac{h}{2}[\rho_\downarrow-\rho_\uparrow]
-\frac{1}{2}\rho_{\uparrow\downarrow} \overline{\Gamma}_{+}(x),
\label{m04} \\ && \frac{ \partial
\rho_{\downarrow\uparrow}}{\partial t}= -i\left[{ H}_{v},
\rho_{\downarrow\uparrow}\right]-i\frac{h}{2}[\rho_\downarrow-\rho_\uparrow]
-\frac{1}{2}\overline{\Gamma}_{+}(x)\rho_{\uparrow\downarrow},
\label{m05}
\\ && \frac{\partial \rho_{2}}{\partial t}= -i\left[{ H}_{v}-x d,
\rho_{2}\right]\nonumber\label{m06}\\
&&\hspace{1.1cm}-\left\{\overline{\Gamma}_R(x),\rho_2\right\}/2
+\sqrt{\overline{\Gamma}_L(x)}\rho_\uparrow\sqrt{\overline{\Gamma}_L(x)},
\end{eqnarray}
where
$\overline{\Gamma}_{+}(x)=\overline{\Gamma}_L(x)+\overline{\Gamma}_R(x)$.
The set of equations (\ref{m01})-(\ref{m06}) is derived in the high
bias voltage limit: $eV/2-\varepsilon_0-U\gg\hbar\omega_0$. In
general, the problem can be solved in two limits with and without
the Coulomb blockade regime. In the Coulomb blockade regime the second
electron can not tunnel onto the quantum dot due to Coulomb repulsion.
Hence the probability for double occupancy $\rho_2\rightarrow0$. 
First, we focus on the case without Coulomb blockade.

Here we repeat the analysis scheme for the evolution of the stationary
solution $\overline{W}_{+}(A)$ for the probability of the shuttle to
vibrate with an amplitude $A$. Expanding the function
$\overline{D}_1(A)$ around $A=0$ one can get the condition for the
shuttle instability
$\gamma<\gamma_{thr}=\Gamma(2h^2d)/\lambda(h^2+\Gamma^2)$. As in the
case of spinless electron, the function $W_{+}$ has a maximum at $A=0$
(stable point) when dissipation rate $\gamma$ is above the threshold
value. In the opposite case the vibrational ground state is
unstable.

The positive bounded function $\beta_0(A,
h)=(2\overline{D}_1(A)-\gamma)\lambda/d$ has only one maximum and
monotonically decreases for large $A$. In \cite{5} it was shown that
if $h<\sqrt{3}\Gamma$, the function $\beta_0$ has a maximum at
$A=0$, while for $h>\sqrt{3}\Gamma$, this function has a minimum at
$A=0$. The structure of the function $\beta_0$ determines the
behavior of the system in the parameter space $d-h$ (or
$\gamma-h$). There are several areas or phases. In the first phase
(vibronic), defined by $d/\gamma\lambda<1/h[max \beta_0(A)]$, the
system is in the lowest vibrational state ($A=0$ is a stable point).
The shuttle phase is developed when $\gamma<\gamma_{thr}$ and there is
only one stable point at $A\neq0$. The third phase is the mixed
phase. It appears because the two above phases become unstable  if $h$
exceeds the critical value $\sqrt{3}\Gamma$.

In the Coulomb blockade regime the same analysis gives that
$\overline{D}_1(A)$ is positive for all values of $h$ if
$\Gamma<4/3$. On the other hand, if $\Gamma>4/3$, there is a range
of magnetic field strenghts where a shuttle instability does not
occur. In particular, when $\Gamma\gg1$ this interval is
$0<h<\Gamma/\sqrt{2}$. This implies that in the adiabatic regime of
charge transport ($\Gamma\gg1$)  in weak magnetic field there is no
instability and the electrically driven electron shuttle is realized
only in strong magnetic fields.

\subsection{Electron Shuttle Based on Electron Spin}

In the previous subsection we studied the shuttle instability in the
case of an electromechanical coupling between the quantum dot and
the leads. In the Coulomb blockade regime a shuttle instability
appears if an external magnetic field $h$ exceeds the critical value
$h_{\text{cr}}=\sqrt{3}\Gamma$. Here we will study the shuttle
instability in the case when the interaction between the dot and the
leads is due to a magnetic (exchange) coupling \cite{kulinichprl}.

The Hamiltonian of the system is similar to the one considered in
Section III. C. The only difference is that the quantum dot
Hamiltonian reads
\begin{eqnarray} \label{Hd}
&&{ H}_{dot}=\varepsilon_0(a^\dag_\uparrow
a_\uparrow+a^\dag_\downarrow
a_\downarrow)\nonumber\\&&\hspace{0.9cm}-J_L(x)(a^\dag_\uparrow
a_\uparrow-a^\dag_\downarrow a_\downarrow)-J_R(x)(a^\dag_\downarrow
a_\downarrow-a^\dag_\uparrow
a_\uparrow)\nonumber\\&&\hspace{0.9cm}-\frac{g\mu
H}{2}(a^\dag_\uparrow a_\downarrow +a^\dag_\downarrow
a_\uparrow)-Ua^\dag_\uparrow a^\dag_\downarrow a_\uparrow
a_\downarrow\,.
\end{eqnarray}
In what follows we will consider the symmetrical case,
$J_R(x)=J_L(-x)$ and restrict ourselves to the Coulomb blockade
regime, $U\sim e^2/2C>\mid eV/2-\varepsilon_0\mid$.

Following Ref.\cite{5} one gets equations of motion for the reduced
density matrix  operators $\rho_{0}\equiv
\langle0\mid\rho\mid0\rangle$, $\rho_{\uparrow} \equiv
\langle\uparrow\mid\rho\mid\uparrow\rangle$,
$\rho_{\downarrow}\equiv
\langle\downarrow\mid\rho\mid\downarrow\rangle$, and
$\rho_{\uparrow\downarrow}\equiv
\langle\uparrow\mid\rho\mid\downarrow\rangle$:
\begin{eqnarray}
&&\frac{\partial \rho_0}{\partial t}=-i\left[{ H}_{v},
\rho_0\right]\nonumber\label{01}\\
&&\hspace{1.1cm}-\left\{\Gamma_L(x),\rho_0\right\}/2
+\sqrt{\Gamma_R(x)}\rho_\downarrow\sqrt{\Gamma_R(x)},\\
&&\frac{\partial \rho_\uparrow}{\partial t}=-i\left[{
H}_{v},\rho_\uparrow\right]+i\left[J(x),
\rho_\uparrow\right]\nonumber\\&&\hspace{1.1cm}- i
h\left(\rho_{\uparrow\downarrow}
-\rho^{\dag}_{\uparrow\downarrow}\right)/2+\sqrt{\Gamma_L(x)}\rho_0
\sqrt{\Gamma_L(x)},\\&& \frac{\partial \rho_\downarrow}{\partial
t}=-i\left[{
H}_{v},\rho_\downarrow\right]-i\left[J(x),\rho_\downarrow\right]
\nonumber\\&&\hspace{1.1cm} +i
h\left(\rho_{\uparrow\downarrow}-\rho^{\dag}_{\uparrow\downarrow}
\right)/2 - \left\{\Gamma_R(x),\rho_\downarrow\right\}/2,
\\&& \frac{\partial \rho_{\uparrow\downarrow}}{\partial t}=
-i\left[{ H}_{v},
\rho_{\uparrow\downarrow}\right]+i\left\{J(x),\rho_{\uparrow
\downarrow}\right\}\nonumber\\&&\hspace{1.3cm} +i h
\left(\rho_\downarrow-\rho_\uparrow\right)/2
-\rho_{\uparrow\downarrow} \Gamma_R(x)/2 \label{02}
\end{eqnarray}
In Eqs.~(\ref{01})-(\ref{02}) $\,\Gamma_j(x)=\Gamma
\text{exp}(j2x/\lambda)$ and $J(x)=J_L(x)-J_R(x)$. In what follows
we assume a linear $x$-dependence of $J(x)$: $J(x)\simeq -\alpha
x+...,\,\alpha=2J'_{R}(0)>0$.

The difference between our operator equations and the corresponding
equations in Ref. \cite{5} (rewritten for the Coulomb blockade case)
is the appearance of terms induced by the coordinate-dependent
exchange interaction $J(x)$. These appear in
Eqs.~(\ref{01})-(\ref{02}) as a commutator term for
$\rho_{\uparrow}$ and $\rho_{\downarrow}$ and as an anti-commutator
term for $\rho_{\uparrow\downarrow}$. In contrast to the
electrically driven shuttle, the driving force in our case is
strongly connected to the spin dynamics, which results in a
completely different dependence of the shuttle behavior on magnetic
field.

Both linear and nonlinear regimes of the shuttling dynamics can be
conveniently analyzed by using the Wigner function representation of
the density operators \cite{novotny}. This approach allows one to
calculate the Wigner distribution function $W_\rho(x,p)$ for the
vibrational degree of freedom to lowest order in the small
parameters $\alpha$ and $1/\lambda$ for small (compared to $\lambda$)
shuttle vibration amplitudes $A$.  The relevant Wigner function,
$W_\Sigma^{(0)}(A)$, averaged over the shuttle phase $\varphi$
($x=A\sin\varphi$), solves the stationary Fokker-Planck equation as
in Eq. (\ref{FPeq}) with drift- and diffusion coefficients containing
the factors
\begin{eqnarray}\label{D1}
&&D_1=\frac{\alpha}{\lambda}\frac{h^2\Gamma^3}{\Gamma^2+3h^2}
\frac{3\Gamma^2+3-h^{2}}{Q_0(\Gamma,h)}  \\
&&D_0=\frac{h^2\Gamma}{\Gamma^2+3h^2}
\left[\frac{\alpha^{2}Q_1(\Gamma,h)+\lambda^{-2}Q_0(\Gamma,h)}{2Q_0(\Gamma,h)}\right]
\label{D0}
\end{eqnarray}
respectively, where
\begin{eqnarray}\label{Q0}
Q_0(\Gamma,h)=\left(1-h^2-2\Gamma^2\right)^2+\frac{\Gamma^2}{4}
\left(\Gamma^2+3h^2-5\right)^2,\hspace{0.3cm}\\\label{Q1}
Q_1(\Gamma,h)=\left(1+\frac{9\Gamma^2}{4}\right)\left(1+h^2+
2\Gamma^2\right)-\frac{5\Gamma^4}{4}.\hspace{1.0cm}
\end{eqnarray}
In Eqs.~(\ref{D1})-(\ref{Q1}) all energies are normalized with
respect to the energy quantum $\hbar\omega$ of the mechanical
vibrations: $\hbar\omega\rightarrow 1$, $g\mu
H/\hbar\omega\rightarrow h$, $J(x)/\hbar\omega\rightarrow J(x)$,
$\Gamma_j(x)/\omega\rightarrow\Gamma_j(x)$
[$\hbar\Gamma_j(x)=2\pi\nu\mid T_j(x)\mid^2$ are partial level
widths].

For $A\ll 1$ the solution of Eq.~(\ref{FPeq})  takes the form of a
Boltzmann distribution function, $W_\Sigma^{(0)}\sim\exp(-\beta
{\cal E})$, where ${\cal E}=A^2/2$ is the dot's vibrational energy
and $1/\beta$, where
\begin{equation}\label{241}
\beta=\left(\frac{2\alpha\Gamma^2}{\lambda}\right)\frac{h^2-3\Gamma^2-3}
{\alpha^2Q_1(\Gamma,h)+\lambda^{-2}Q_0(\Gamma,h)}\,,
\end{equation}
is an effective temperature. Since the functions $Q_0$ and $Q_1$ are
positive, the sign of the effective temperature is determined by the
relation between magnetic field, level width and vibration quantum.
In particular the effective temperature is negative at small
magnetic fields, $|H| < H_c$, where (reverting to dimensional
variables) $g \mu H_c=\hbar\sqrt{3\left(\Gamma^2+\omega^2\right)}$.

A negative $\beta$ implies that the static state of the dot ($A=0$)
is unstable and that a shuttling regime of charge transport ($A\ne
0$) is realized. It is interesting to note that $\beta$ is finite
even as $h\rightarrow 0$. This apparent paradox may be resolved by
considering the Fokker-Plank equation in its time-dependent form and
noting that the rate of change of the oscillation amplitude at the
instability is defined by the coefficient $D_1$. This coefficient
scales as $D_1(h)\propto h^2$ as $h\rightarrow 0$ and therefore the
shuttle phase is only realized formally after an infinitely long
time in this limit. As a function of magnetic field $D_1$ has a
maximum, $D^{max}_{1}=0.6 (\alpha/\lambda)\Gamma^{-1}$, at $h_{opt}=
0.4 \Gamma$. Therefore, optimal magnetic fields are in the range
$0.1-1$~T if $\hbar\Gamma =10-100$~$\mu$eV. For high magnetic
fields, $|H|> H_c$, there is no shuttling regime (at least not with
a small vibration amplitude, $A\ll 1$) and the vibronic regime,
corresponding to small  fluctuations of the quantum dot around its
equilibrium position, is stable.

\begin{figure}
\centering
\includegraphics[width=0.85\columnwidth]{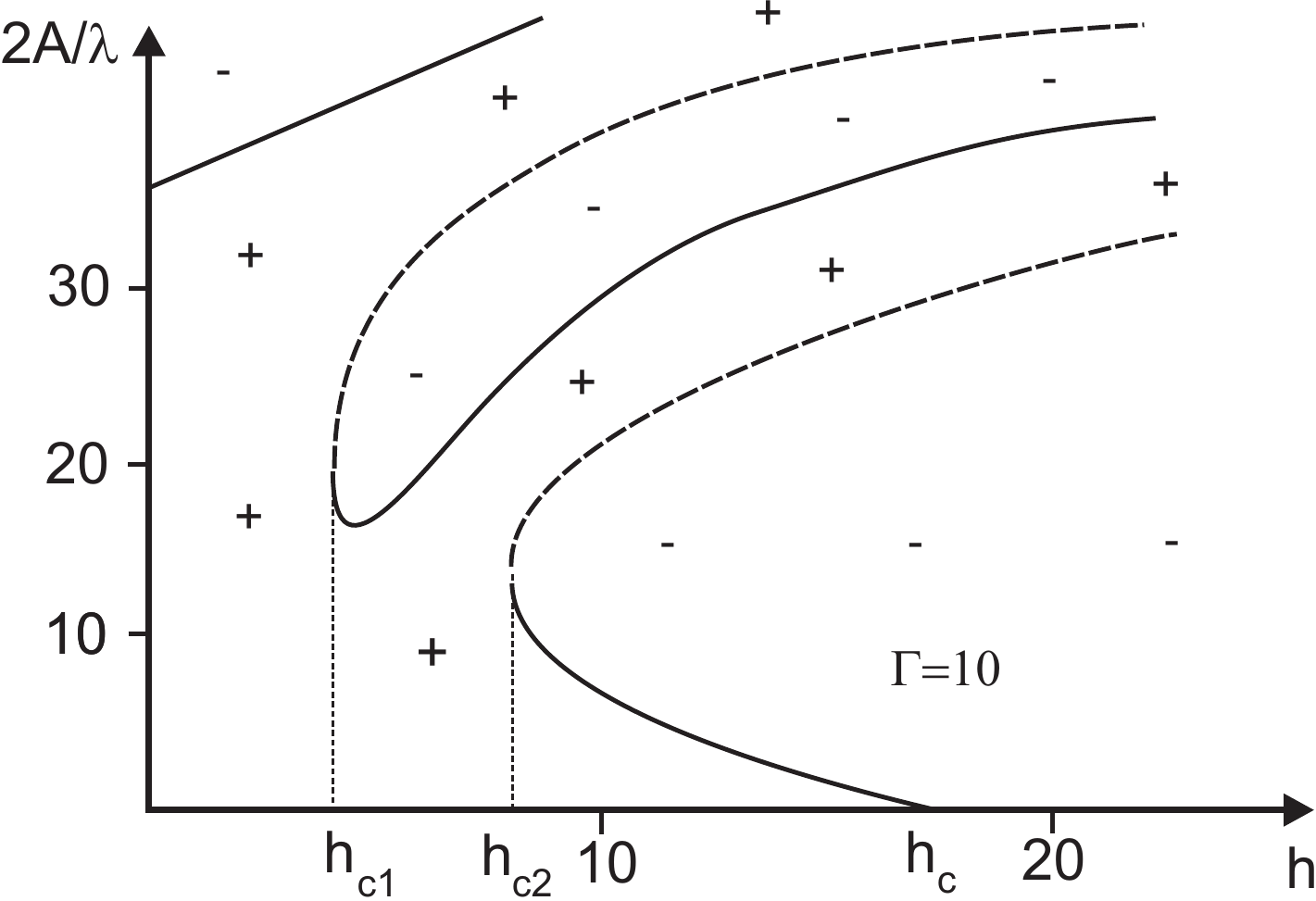}
\caption{Regions of positive and negative values of the increment
coefficient $D_1(A,h)$ for $\Gamma=10$. Solid (dashed) lines
indicate where the Wigner distribution function for the oscillation
amplitude $A$ has a local maximum (minimum) and hence where the
stationary state [$D_1(A,h)=0]$ is stable (unstable) with respect to
small perturbations. }\label{Fig2}
\end{figure}

The amplitude of the shuttle vibrations that develop as the result
of an instability is still described by Eq.~(\ref{FPeq}) for the
Wigner distribution function. However, for large amplitudes,
$A\gtrsim 1$, the drift- and diffusion coefficients $A^{2}D_1$ and
$AD_0$ can no longer be evaluated analytically. Fortunately, it is
sufficient to know the amplitude- and magnetic field dependence of
$D_1$ for a qualitative analysis. This is because a positive value
of the drift coefficient means that energy is pumped into the dot
vibrations, while a negative value corresponds to damping (cooling)
of the vibrations. Therefore, magnetic fields for which $D_1(A)=0$
and $D_1'(A)<0$ correspond to a stable stationary state of the dot
and a local maximum of the Wigner function. Based on this picture
one concludes (see Fig.~\ref{Fig2}) that at low magnetic fields,
$h<h_{c1}$, a shuttling regime with a large vibration amplitude is
realized, while at high magnetic fields, $h>h_{c1}$ the situation is
more complicated.  Here one of two ($h_{c1}<h<h_{c2}$; $h>h_c$) or
three ($h_{c2}<h<h_{c}$) shuttling regimes with different amplitudes
can be stable depending on the initial conditions . If the dot is
initially in the static state ($A=0$) a stable shuttle regime only
appears for $h<h_c$ as already mentioned.

Thus the magnetic shuttle device acts in "opposite" way as compared
to electromechanical one.  A particularly transparent picture of how
spintro-mechanics affects shuttle vibrations emerges in the limit of
weak magnetic field $H$ and large electron tunnelling rate
$\Gamma_{S(D)}$ between dot and source- and drain electrodes. In
order to explore this limit, where $\Gamma_{S}\gg\omega\gg(\mu
H/\hbar)^{2}/\Gamma_{D}$ and $\omega/2\pi$ is the natural vibration
frequency of the dot, we focus first on the total work done by the
exchange force $\mathfrak{F}$ as the dot vibrates under the
influence of an elastic force only. In the absence of an external
magnetic field the dot is in this case occupied by a spin-up
electron emanating from the source electrode. This spin is a
constant of motion and hence no electrical current through the
device is possible since only spin-down states are available in the
drain electrode. During the oscillatory motion of the dot the
exchange force is therefore always directed towards the source
electrode while its magnitude only depends on the position of the
dot, $\mathfrak{F}=\mathfrak{F}_{0}(x)$. As a result, no net work is
done by the exchange force on the dot. This is because contributions
are positive or negative depending on the direction of the dot's
motion and cancel when summed over one oscillation period. A finite
amount of work can only be done if the exchange force deviates from
$\mathfrak{F}_{0}(x)$ as a result of spin flip processes induced by
the external magnetic field. Such a deviation can be viewed as an
additional random force $\mathfrak{F}_{H}$ that acts in the opposite
direction to $\mathfrak{F}_{0}(x)$. In the limit of large tunneling
rate, $\Gamma\gg\mu H/\hbar$, and small vibration amplitude a spin
flip occurs  with a probability $\propto(\mu
H/\hbar)^{2}/(\omega\Gamma_{D})$ during one oscillation period and
is instantly accompanied by the tunneling of the dot electron into
the drain electrode, thereby triggering the force
$\mathfrak{F}_{H}$. The duration of this force is determined by the
time $\delta t\sim 1/\Gamma_{S}(x(t))$ it takes for the spin of the
dot to be ``restored" by another electron tunneling from the source
electrode.

The spin-flip induced random force $\mathfrak{F}_{H}
=-\mathfrak{F}_{0}(x)$  is always directed towards the drain
electrode. Hence, its effect depends on the dot's direction of
motion: as the dot moves away from the source electrode it will be
accelerated, while as it moves towards the source it will be
decelerated. Since a spin-flip may occur at any point on the
trajectory one needs to average over different spin-flip positions
in order to calculate the net work done on the dot. The result,
which depends on the competition between the effect of spin flips
that occur at the same position but with the dot moving in opposite
directions, is nonzero because $\delta t$ is different in the two
cases. As the dot moves away from the source electrode the tunneling
rate to this electrode will decrease while as the dot moves towards
the source it will increase. This means that the duration of
spin-flip induced acceleration will prevail  over the one for
deceleration. As a result, in weak magnetic fields, the dot will
accelerate with time and one can expect a spintro-mechanical
shuttle instability in this limit.

The situation is qualitatively different in the opposite limit of
strong magnetic fields, where $\Gamma\ll\mu H/\hbar$ and the spin
rotation frequency therefore greatly exceeds the tunneling rates. In
this case the quick precession of the electron spin in the dot
averages the exchange force to zero if one neglects the small
effects of electron tunneling to and from the dot. If one takes
corrections due to tunnelling into account (having in mind that the
source electrode only supplies spin-up electrons) one comes to the
conclusion that the average spin on the dot will be directed
upwards. This results in a net spintro-mechanical force in the
direction opposite to that of the net force occurring in a weak
magnetic field limit. As a result, in strong magnetic fields one
expects on the average a deceleration of the dot. Therefore, there
will be no shuttle instability for such magnetic fields.

As we have discussed above spin-flip assisted electron tunnelling
from source to dot to drain in our device results in a magnetic
exchange force that attracts the dot to the source electrode. It is
interesting to note that this is contrary to the effect of the
Coulomb force in the same device. Indeed, since the Coulomb force
depends on the electric charge of the dot it repels the dot from the
source electrode. Hence, while the dot is empty as the result of a
spin-flip assisted tunneling event from dot to drain, an ``extra"
{\em attractive} Coulomb force $\mathfrak{F}_{Q}$ is active. An
analysis fully analogous with our previous analysis of the ``extra"
{\em repulsive} magnetic exchange force $\mathfrak{F}_{H}$  leads to
the conclusion that the effect of the Coulomb force will be just the
opposite to that of the exchange force. This means that in the
Coulomb blockade regime in the limit of weak magnetic field there is
no shuttle instability, while in strong magnetic fields electron
shuttling occurs. As was shown the detailed analysis confirms these
predictions.

\subsection{Mechanically assisted magnetic coupling between
nanomagnets}

The mechanical force caused by the exchange interaction represents
only one effect of the coupling of magnetic and mechanical degrees of
freedom in magnetic nanoelectromechanical device. A complementary
effect is the of mechanical transportation of magnetization, which
we are going to discuss in this subsection.

\begin{figure}
\vspace{-0.cm} \includegraphics[width=7cm]{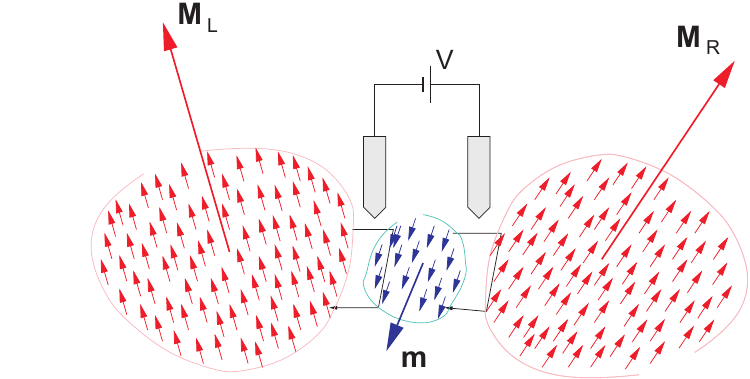} \vspace*{-0.cm}
\caption{Single-domain magnetic grains with magnetic moments
$\textbf{M}_L$ and $\textbf{M}_R$ are coupled via a magnetic cluster
with magnetic moment $\textbf{m}$, the latter being separated from
the grains by insulating layers. The gate electrodes induce an ac
electric field, concentrated in the insulating regions. This field,
by controlling the heights of the tunnel barriers, affects the
exchange magnetic coupling between different components of the
system. Reprinted with permission from \cite{8}, L. Y. Gorelik et
al., {Phys. Rev. Lett.} {\bf 91},
 088301 (2003). $\copyright$ 2003, American Physical Society.}
\end{figure}

In the magnetic shuttle device presented in Fig.~6, a ferromagnetic
dot with total magnetic moment $\textbf{m}$ is able to move between
two magnetic leads, which have total magnetization
$\textbf{M}_{L,R}$. Such a device was suggested in
Ref.~\onlinecite{8} in order to consider the magnetic coupling
between the leads (which in their turn can be small magnets or
nanomagnets) produced by a ferromagnetic shuttle. It is worth to
point out that the phenomenon we are going to discuss here has
nothing to do with transferring electric charge in the device and it
is valid also for a device made of nonconducting material. The main
effect, which will be in the focus of our attention, is the exchange
interaction between the ferromagnetic shuttle (dot) and the magnetic
leads. This interaction decays exponentially when the dot moves away
from a lead and hence it is only important when the dot is close to
one of the leads. During the periodic back-and-forth motion of the
dot this happens during short time intervals near the turning points
of the mechanical motion. An exchange interaction between the
magnetizations of the dot and a lead results in a rotation of these
two magnetization vectors in such a way that the vector sum is
conserved. This is why the result of this rotation can be viewed as
a transfer of some magnetization $\Delta\textbf{m}$ from one
ferromagnet to the other. As a result the magnetization of the dot
experiences some rotation around a certain axis. The total angle
$\phi$ of the rotation accumulated during the time when the dot is
magnetically coupled to the lead is an essential parameter which
depends on the mechanical and magnetic characteristics of the
device. The continuation of the mechanical motion breaks the
magnetic coupling of the dot with the first lead but later, as the
dot
approaches
the other magnetic lead an exchange coupling is established with
this second lead with the result that
magnetization which is ``loaded" on the dot from the first lead is
"transferred" to the this second lead.
This is how the transfer of magnetization from one magnetic lead to
another is induced mechanically. The transfer creates an
effective coupling between the magnetizations of the two leads. Such
a non-equilibrium coupling can be efficiently tuned by controlling
the mechanics of the shuttle device. It is particularly interesting
that the sign of the resulting magnetic interaction is determined by
the sign of $\cos (\phi/2)$. Therefore, the mechanically mediated
magnetic interaction can be changed from ferromagnetic to
anti-ferromagnetic by changing the amplitude and the frequency of
mechanical vibrations \cite{8}.

\section{Resonance spin-scattering effects. Spin shuttle as a "mobile quantum impurity".}

{ Many-particle effects add additional dimension to the shuttling
phenomena. These effects accompany electronic tunneling between the
gate electrodes and the moving nanoisland. The common source of
many-particle effects is the so called "orthogonality catastrophe"
related to multiple creation of electron-hole pairs both with
parallel and antiparallel spins \cite{Mahan67,Anderson67} as a
response of electronic gas in the leads to single electron
tunneling. The second-order cotunneling processes under strong
Coulomb blockade result in effective indirect exchange between the
shuttle and the leads. This exchange is the source of strong
scattering and the many-particle reconstruction of the electron
ensemble in the leads  known as the Kondo effect. Various
manifestations of the Kondo effect in shuttling are reviewed in this
section.}

\begin{figure}[t]
\vspace*{0mm}
\includegraphics[angle=0,width=\columnwidth]{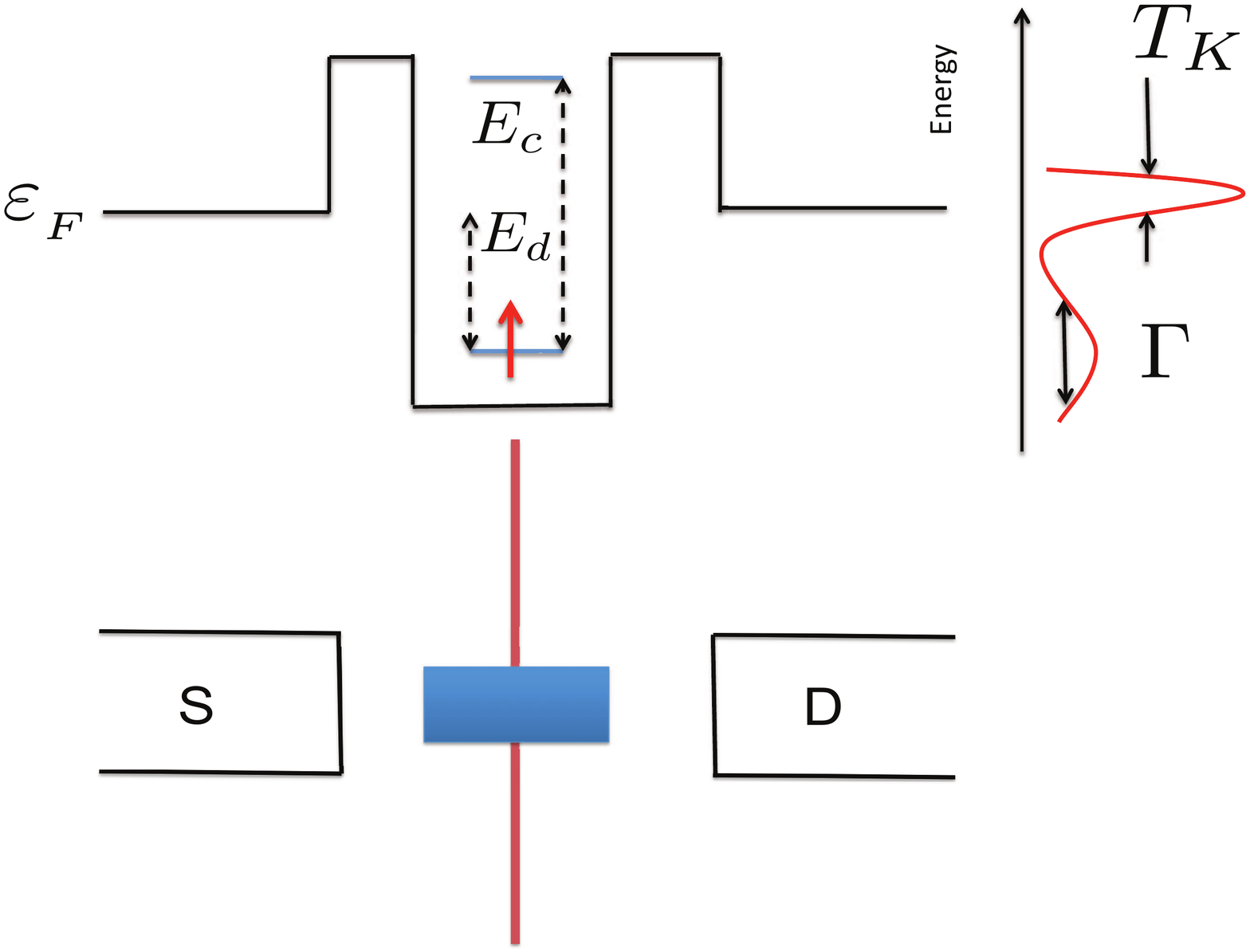}
\vspace*{-2mm} \caption{  Nanomechanical resonator with spin as a
``mobile quantum impurity".} \label{f.1}
\end{figure}

The Kondo effect in electron tunneling close to the unitarity limit
manifests itself as a sharp zero bias anomaly in the low-temperature
tunneling conductance. Many-particle interactions renormalize the
electron spectrum enabling  "Abrikosov-Suhl resonances"  both for
odd \cite{gogo} and even \cite{Pust00,Kikoin01} electron
occupations. In the latter case the resonance is caused by the
singlet-triplet crossover in the ground state (see \cite{glasko} for
a review). In the simplest case of odd occupancy a cartoon of a
quantum well and a schematic Density of States (DoS) is shown in
Fig.~\ref{f.1}. For simplicity we consider a case when the dot is
occupied by one electron (as in a SET transistor). { The
corresponding electronic level in the dot is located at an energy
$-E_d$, deep beyond the Fermi level of the leads ($\epsilon_F$). The
dot is in the Coulomb blockade regime, and the corresponding
charging energy is denoted as $E_C$. The Abrikosov-Suhl resonance
\cite{Abrikosov, Suhl, Hewson} at $\epsilon_F$ arises due to
multiple spin flip scattering, so that the narrow peak in the DoS is
related mainly to the spin degrees of freedom} (see Fig.~\ref{f.1},
upper right panel). The width of this resonance is defined by the
unique energy scale, the Kondo temperature
 $T_K$, which determines all thermodynamic and transport properties of the
SET device through a one-parametric scaling \cite{Hewson}.
The Breit - Wigner (BW) width $\Gamma$ of the dot level associated
with the tunneling of dot electrons to the continuum of levels in
the leads, is assumed to be smaller
than the charging energy $E_C$, providing a condition for nearly
integer valency regime.

Building on an analogy with the shuttling experiments of Refs.
\onlinecite{Kot2004}
 and \onlinecite{erbe}, let us consider a device where an isolated
nanomachined island oscillates between two electrodes. The applied
voltage is assumed low enough so that the field emission of many
electrons, which was the main mechanism of tunneling in those
experiments, can be neglected. We emphasize that the characteristic
de Broglie wave length associated with the dot should be much
shorter than typical displacements allowing thus for a classical
treatment of the mechanical motion of the nano-particle. The
condition $\hbar\omega_0\ll k_B T_K$,  necessary to eliminate
decoherence effects, requires for e.g. planar quantum dots with the
Kondo temperature $T_K\gtrsim 100$~mK, the condition
$\omega_0\lesssim 1$~GHz for oscillation frequencies to hold; this
frequency range is experimentally feasible \cite{erbe,Kot2004}. The
shuttling island is then to be considered as a ``mobile quantum
impurity", and transport experiments will detect the influence of
mechanical motion on the differential conductance. If the dot is
small enough, then the Coulomb blockade guarantees the single
electron tunneling or cotunneling regime, which is necessary for the
realization of the Kondo effect \cite{GR,glasko}.

{ The above configuration is illustrated in the lower panel of Fig.
\ref{f.1}: the shuttle of nanoscale size is mounted at the tight
string. Its harmonic oscillations are induced
 by external elastic force.
Unlike the conventional resonance case (the  resonance level belongs
not to the moving shuttle but develops as a many-body peak at the
Fermi level of the leads. }
 When the shuttle moves between source (S) and drain
(D) (see the lower panel of Fig. \ref{f.1}), both the energy $E_d$
and the
width $\Gamma$ acquire a time dependence. This time dependence
results in a coupling between mechanical, electronic and spin
degrees of freedom. If a source-drain voltage $V_{sd}$ is small
enough ($eV_{sd}\ll k_B T_K$) the charge degree of freedom of the
shuttle is frozen out while spin flips play a very important role in
co-tunneling processes. { Namely, the Abrikosov-Suhl resonance is
viewed as a time-dependent Kondo cloud built up from conduction
electrons in the leads dynamically screening moving spin localized
at the shuttle.} Since the electrons in the cloud contain
information about the same impurity, they are mutually correlated.
Thus, NEM  providing a coupling between mechanical and electronic
degrees of freedom introduces a powerful tool for manipulation and
control of the Kondo cloud {induced by the spin scattering} and
gives a very promising and efficient mechanism for electromechanical
transduction on the nanometer length scale.

 Cotunneling
is accompanied by a change of spin projection in the process of
charging/discharging of the shuttle and therefore is closely related
to the spin/charge pumping problem \cite{brouw98}.

A generic Hamiltonian for describing the resonance spin-scattering
effects is given by the same Anderson model as above,
\begin{eqnarray}
&&H_0 =\sum_{k,\alpha}
\varepsilon_{k\sigma,\alpha}a^\dagger_{k\sigma,\alpha}a_{k\sigma,\alpha}
+\sum_{i\sigma}[E_{d}-e Ex] d^{\dagger}_{i\sigma} d_{i\sigma} + E_C
n^2 \nonumber
\\
&& H_{tunnel}=
\sum_{ik\sigma,\alpha}T^{(i)}_{\alpha}(x)[a^\dagger_{k\sigma,\alpha}d_{i\sigma}
+ H.c],\label{ham1}
\end{eqnarray}
where  ${ E}$ is the electric field between the leads. The
tunnelling matrix element depends exponentially on the ratio of the
time-dependent displacement $x(t)$ and the electronic tunnelling
length $\lambda$, see Eq. (\ref{Spintro Tunn}).  The time-dependent
Kondo Hamiltonian for slowly moving shattle can be obtained by
applying a time-dependent Schrieffer-Wolff transformation
\cite{SW,KNG00}:
\begin{eqnarray}
H_K=\sum_{k\alpha\sigma, k'\alpha'\sigma'}{\cal
J}_{\alpha\alpha'}(t)[\vec{\sigma}_{\sigma\sigma'}\vec{S}+
\frac{1}{4}\delta_{\sigma\sigma'}]a^\dagger_{k\sigma,\alpha}a_{k'\sigma',\alpha'}
\label{ham2}
\end{eqnarray}
where ${\cal
J}_{\alpha,\alpha'}(t)=\sqrt{\Gamma_{\alpha}(t)\Gamma_{\alpha'}(t)/(\pi\rho_0
E_d(t))}$ and $\vec
S=\frac{1}{2}d_\sigma^\dagger\vec\sigma_{\sigma\sigma'}
d_{\sigma'}$, $\Gamma_\alpha(t)=2\pi \rho_0 |T_\alpha(x(t))|^2$ are
level widths due to tunneling to the left and right leads.

As long as the nano-particle is not subject to an external
time-dependent electric field, the Kondo temperature is given by
$k_B T_K^0=D_0\exp\left[-(\pi E_C)/(8\Gamma_0)\right]$ (for
simplicity we assumed that $\Gamma_L(0)=\Gamma_R(0)=\Gamma_0$; $D_0$
plays the role of effective bandwidth). As the nano-particle moves
adiabatically, $\hbar \omega_0\ll \Gamma_0$, the decoherence effects
are small provided $\hbar\omega_0\ll k_B T_K^0$.

\begin{figure}[h]
  \includegraphics[width=80mm,angle=0]{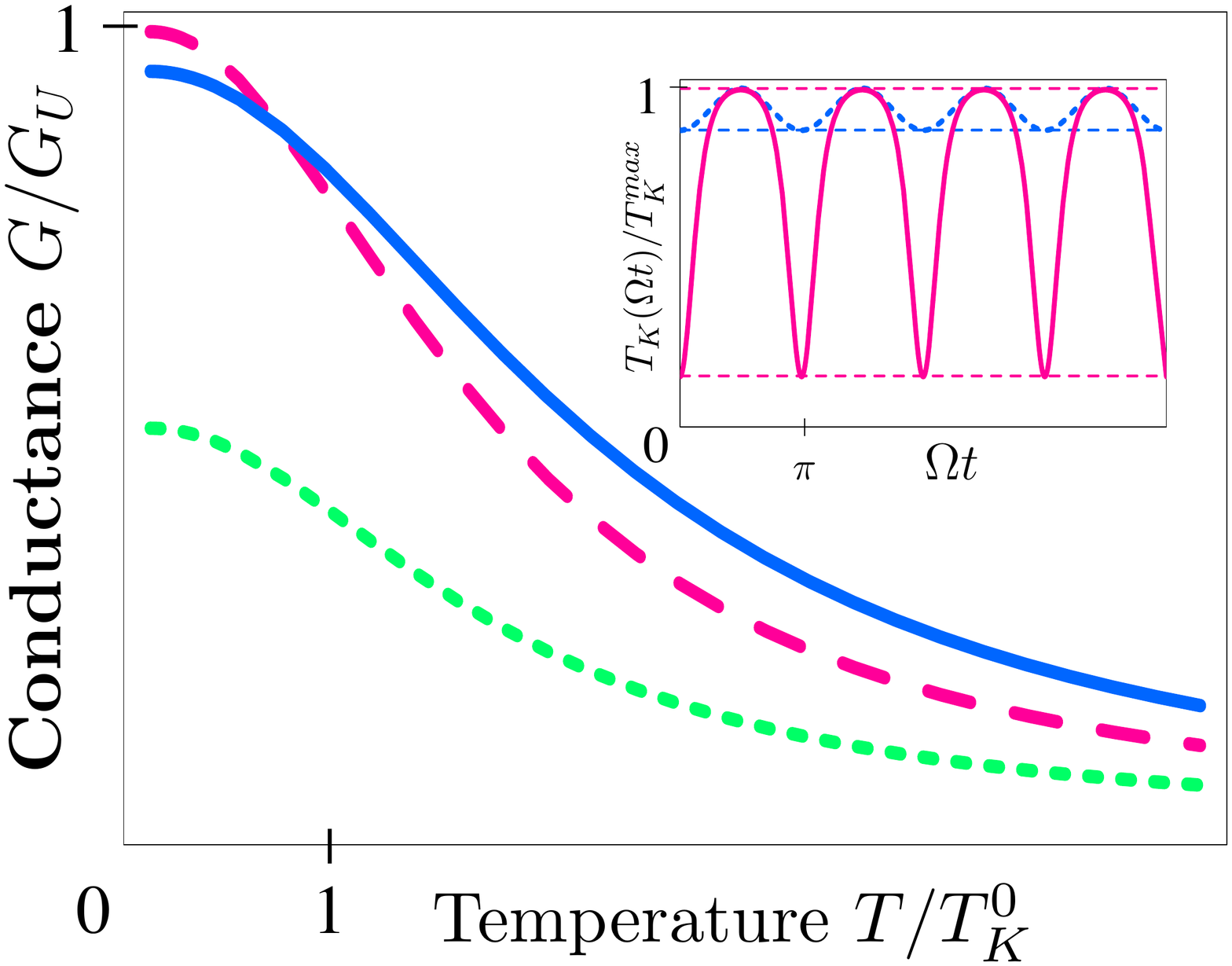}
  \caption{ Differential conductance $G$ of a Kondo shuttle for which $\Gamma_0$$/$$E_C$$=$$0.4$.
  The solid line denotes $G$ for a shuttle with $\Gamma_L$$=$$\Gamma_R$, $A$$=$$\lambda$,
 the dashed line shows $G$ for a static nano-island with $\Gamma_L=\Gamma_R$, $A$$=$$0$,
 the dotted line gives $G$ for $\Gamma_L$$/$$\Gamma_R$$=$$0.5$, $A$$=$$0$.
 The inset shows the  temporal oscillations (here $\Omega\equiv\omega_0$) of $T_K$ for small
 $A$$=$$0.05$$\,\lambda$ (dotted line) and large $A$$=$$2.5$$\,\lambda$ (solid line) shuttling amplitudes.
 Reprinted with permission from \cite{kis06},
M. N. Kiselev {\em et al.}, {Phys. Rev. B} {\bf 74},
 233403 (2006). $\copyright$ 2006, American Physical Society.
  }\label{fig:sh2}
\end{figure}

Let us first assume a temperature regime $T\gg T_K$ (weak coupling).
In this case we can build a perturbation theory controlled by the
small parameter $\rho_0 {\cal J}(t)\ln[D_0/(k_B T)]<1$ assuming time
as an external parameter. The series of perturbation theory can be
summed up by means of a renormalization group procedure
\cite{Hewson,KNG00}. As a result, the Kondo temperature becomes
oscillating in time:
\begin{eqnarray}
k_B T_K(t)=D(t) \exp\left[-\frac{\pi
E_C}{8\Gamma_0\cosh(2x(t)/\lambda)}\right].\label{ftk}
\end{eqnarray}
Neglecting the weak time-dependence of the effective bandwidth
$D(t)\approx D_0$, we arrive at the following expression for the
time-averaged Kondo temperature:
\begin{eqnarray}
\langle T_K\rangle=T_K^0 \bigg\langle \exp\left[\frac{\pi
E_C}{4\Gamma_0}\frac{\sinh^2 (x(t)/\lambda)}{1+2\sinh^2
(x(t)/\lambda)}\right]\bigg\rangle. \label{tkon1}
\end{eqnarray}
Here $\langle$$...$$\rangle$ denotes averaging over the period of
the mechanical oscillation. The expression (\ref{tkon1}) acquires an
especially transparent form when the amplitude of the mechanical
vibrations $A$ is small: $A\lesssim \lambda$. In this case the Kondo
temperature can be written as $\langle T_K \rangle = T_K^0
\exp(-2W)$, with the Debye-Waller-like exponent $W =-\pi E_C \langle
x^2(t)\rangle)/(8\Gamma_0\lambda^2)$, giving rise to the enhancement
of the static Kondo temperature.

The zero bias anomaly (ZBA) in the tunneling conductance is given by
\begin{eqnarray}
G(T) =\frac{3\pi^2}{8}G_0\Bigg\langle\frac{4\Gamma_L(t)\Gamma_R(t)}
{(\Gamma_L(t)+\Gamma_R(t))^2}\frac{1}{[\ln(T/T_K(t))]^2}\Bigg\rangle,
\end{eqnarray}
where $G_0=e^2/h$ is a unitary conductance. Although the central
position of the island is most favorable for the BW resonance
($\Gamma_L =\Gamma_R$), it corresponds to the minimal width of the
Abrikosov-Suhl resonance. The turning points correspond to the
maximum of the Kondo temperature given by the equation (\ref{ftk})
while the system is away from the BW resonance. These two competing
effects lead to the effective enhancement of $G$ at high
temperatures (see Fig. \ref{fig:sh2}).

Summarizing, it was shown in \cite{kis06} that Kondo shuttling in a
NEM-SET device increases the Kondo temperature due to the asymmetry
of coupling at the turning points compared to at the central
position of the island. As a result, the enhancement of the
differential conductance in the weak coupling regime can be
interpreted as a pre-cursor of strong electron-electron correlations
appearing due to formation of the Kondo cloud.

\begin{figure}[t]
\vspace*{-0mm}
\includegraphics[angle=0,width=7cm]{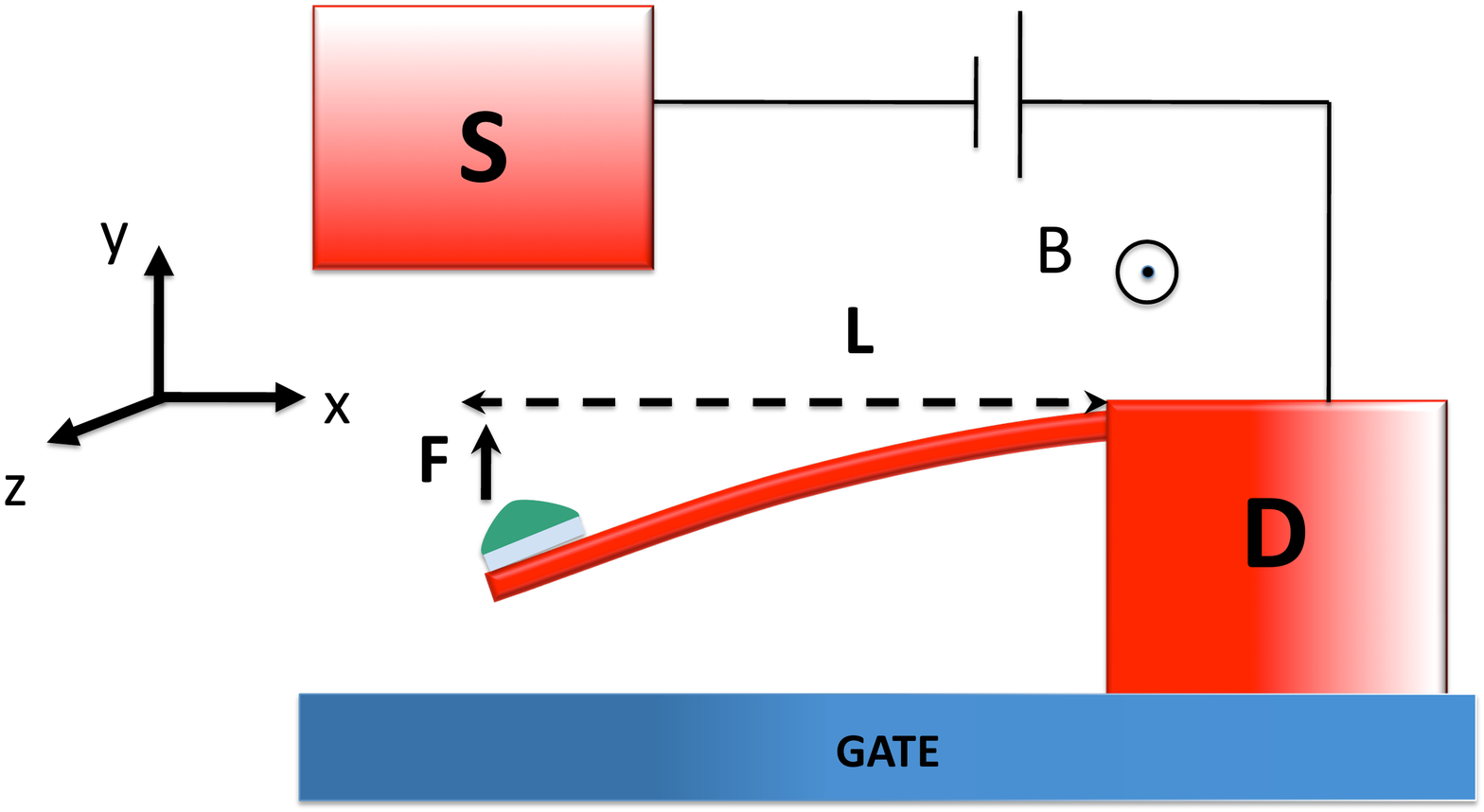}
\vspace*{-0.mm} \caption{Shuttling quantum dot mounted on a moving
metallic pendulum. Magnetic field $B$
  is applied along $z$ axis. $\copyright$ 2013, American Physical Society.} \label{f.pend}
\end{figure}

{ Next we turn to the strong coupling regime, $T\ll T_K$. We
consider this regime for an oscillating cantilever with a nanotip at
its end (Fig. \ref{f.pend}). Then the motion of a shuttle in $y$
direction is described by the Newton equation which we rewrite in a
form
\begin{equation}\label{2.4}
  \ddot{y} + \frac{\omega_0}{Q_0} \dot{y} + \omega^2_0 y =\frac{1}{m} F.
\end{equation}
 where $\omega_0=\sqrt{k/m}$ is the oscillator frequency of free cantilever,
$Q_0$ is the quality factor. $F$ is the Lorentz force acting on
moving cantilever in perpendicular magnetic field
\begin{equation}\label{4.6}
\vec F = L\cdot\vec I\times \vec B =(0,F,0).
\end{equation}
Here $L$ is the length of the cantilever. $\vec I$ is the current
through the system.}

{ In this configuration the Kondo cloud induced by spin scattering
is formed both in the immovable part of the setup (drain electrode)
and in the oscillating cantilever. } The current $\vec I$ subject to
a constant source-drain bias $V_{sd}$ can be separated in two parts:
a dc current associated with a time-dependent dc conductance and an
ac current related to the periodic motion of the shuttle. While the
dc current is mostly responsible for the frequency shift, the ac
current gives an access to the dynamics of the Kondo cloud and
provides information about the kinetics of its formation. In order
to evaluate both contributions to the total current we rotate the
electronic states in the leads in such a way that only one
combination of the wave functions is coupled to the quantum
impurity. The cotunneling Hamiltonian may be rationalized by means
of the Glazman-Raikh rotation, parametrized by the angle
$\vartheta_t$ defined by the relation $\tan \vartheta_t =
\sqrt{|\Gamma_R(t)/\Gamma_L(t)|}$.

Both the ac and dc contributions to the current can be calculated by
using Nozi\`ere's Fermi-liquid theory (see \cite{Nozieres74} for
details). The ac contribution, associated with the time dependence
of the Friedel phase $\delta_{\sigma}$ \cite{kis12}, is given by
\begin{equation}\label{3.15}
\bar I_{ac}(t) = \frac{\dot {y}(t)}{\lambda} \frac{e
E_C}{8\Gamma_0}\cdot\frac{e V_{\rm sd}}{k_B T_K(t)}\cdot \frac{
\tanh\left(\frac{2[y(t)-y_0]}{\lambda}\right)}{
\cosh^{2}\left(\frac{2[y(t)-y_0]}{\lambda}\right)}\\
\end{equation}
\noindent ($\exp(4 y_0/\lambda)=\Gamma_R(0)/\Gamma_L(0)$). The
equation (\ref{3.15}) acquires a simple form if we assume that the
size of Kondo cloud $R_K(y(t))=\hbar v_F/(k_B T_K(y(t)))$ where
$v_F$ is a Fermi velocity. According to Nozieres \cite{Nozieres74},
the Friedel phase $\delta_\sigma$ can be Taylor-expanded in the
vicinity of its resonance value $\delta_{0\sigma}=\pi/2$ as
\begin{equation}\label{nearres}
\delta_\sigma(t)=\frac{\pi}{2}+\frac{e V_{\rm sd} R_K(y(t))}{\hbar
v_F}+\frac{g \mu_B (\sigma\cdot B) R_K(y(t))}{\hbar v_F}
\end{equation}
and, therefore, $d (\delta_\uparrow+\delta_\downarrow)/ dt \propto
\dot y \cdot d R_K(y)/d y$. As a result,
\begin{equation}\label{cloud}
\bar I_{ac}(t) = 2 G_0 V_{\rm sd} \frac{\dot y(t)}{v_F}\frac{d
R_K(y)}{d y}.
\end{equation}
Thus, the ac current generated in the device due to the mechanical
motion of the shuttle contains information about spatial variation
of the Kondo cloud.

The "ohmic" dc contribution is fully defined by the adiabatic
time-dependence of the Glazman-Raikh angle
\begin{equation}\label{3.16}
\bar I_{DC}(t) = G_0 V_{sd} \sin^2
2\vartheta_t\sum_\sigma\sin^2\delta_\sigma
\end{equation}
As a result, the ac contribution to the total current can be
considered as a first non-adiabatic correction:
\begin{equation}\label{4.5}
I_{tot} = I_{ad}(y(t))-\dot y \frac{dI_{ad}}{dy}\frac{\hbar \pi
E_C}{16 \Gamma_0 k_B T^{(0)}_K}
\end{equation}
where  $I_{ad}= 2\cdot  G_0\cdot
V_{sd}\cosh^{-2}(2[y(t)-y_0]/\lambda)$ and $T_K^{(0)}$ is the Kondo
temperature at the equilibrium position. The small correction to the
adiabatic current  in (\ref{4.5}) may be considered as a first term
in the  expansion over the small  non adiabatic parameter
\color{black} $\omega_0\tau\ll 1$\color{black}, where $\tau$ is the
retardation time associated with the inertia of the Kondo cloud.
Using such an interpretation one gets $\tau= \hbar \pi E_C/(16
\Gamma_0 k_B  T^{(0)}_K)$.

\begin{figure}[t]
\vspace*{+5mm}
\includegraphics[angle=0,width=7cm]{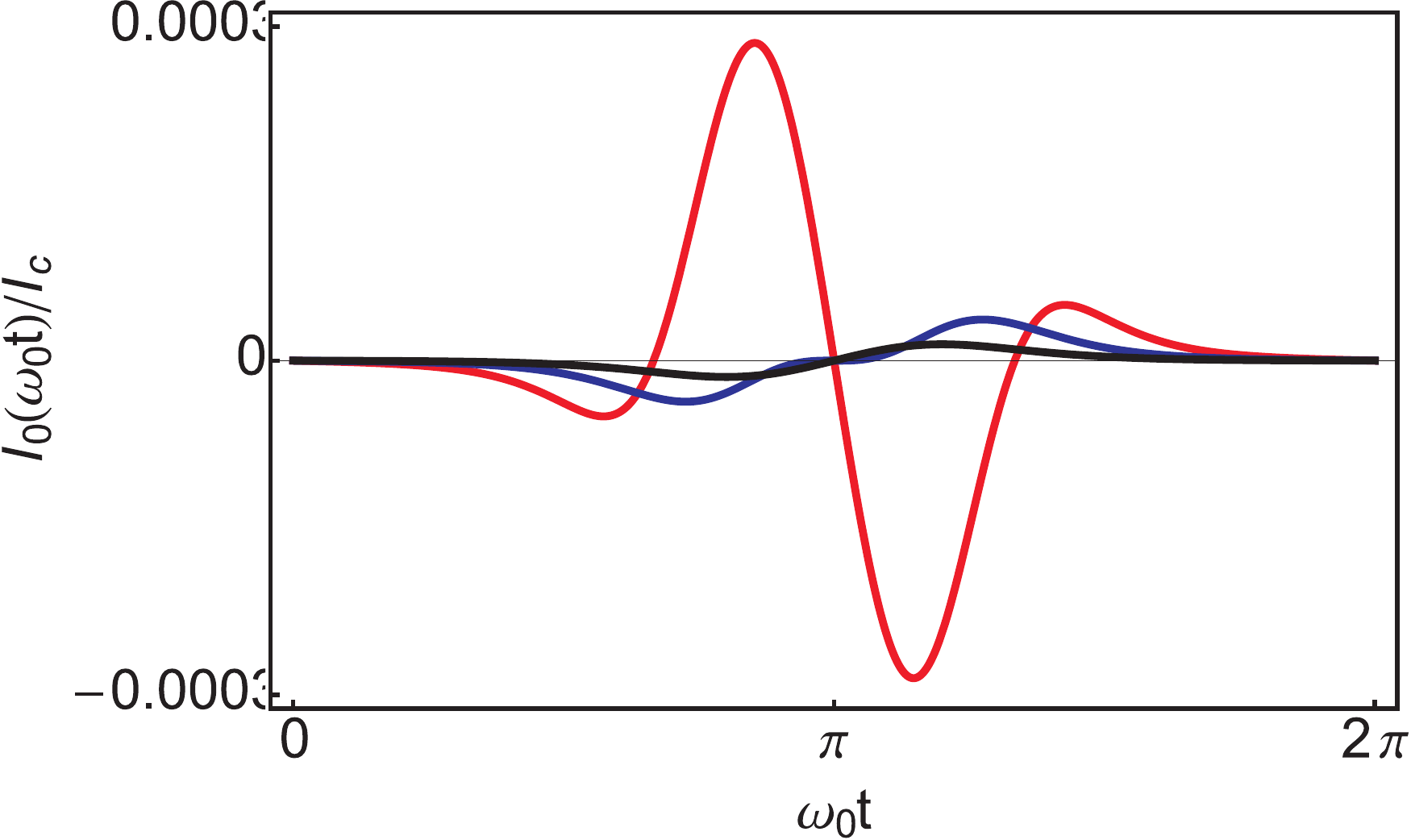}
\vspace*{-0.mm} \caption{Time dependence of the current $I_0$ for
different values of asymmetry parameter $u=x_0/\lambda$. Here red,
blue and black curves correspond to $u=0.5; 1.0; 1.5;$. For all
three curves shuttle oscillates with amplitude $x_{max}=\lambda$,
$\hbar\omega_0/(k_B T_K^{\color{black}min\color{black}})=10^{-3}$,
$|eV_{\rm bias}|/(k_B T_K^{\color{black}min\color{black}})=g\mu_B
B/(k_B T_K^{\color{black}min\color{black}})=0.1$ with $T_K^{(0)}=2
K$, $\lambda/L=10^{-4}$. Reprinted with permission from
\cite{kis12}, M. N. Kiselev {\em et al.} Phys. Rev. Lett. {\bf 110},
066804 (2013). $\copyright$ 2013, American Physical Society.}
\label{ff.1}
\end{figure}

Equation (\ref{4.5}) allows one  to obtain information about the
dynamics of the Kondo clouds from an analysis of an experimental
investigation of the  mechanical vibrations. \color{black} The
retardation time associated with the dynamics of the Kondo cloud is
parametrically large compared with the time of formation of the
Kondo cloud $\tau_K=\hbar/(k_B T_K)$ and can be measured owing to a
small deviation from adiabaticity. \color{black} Also we would like
to emphasize a supersensitivity of the quality factor to a change of
the equilibrium position of the shuttle characterized by the
parameter $u$ (see Fig.~\ref{ff.1}). The influence of strong
coupling between mechanical and electronic degrees of freedom on the
mechanical quality factor has been considered in \cite{kis12}. It
has been shown that both suppression $Q>Q_0$ and enhancement $Q<Q_0$
of the dissipation of nanomechanical vibrations (depending on
external parameters and the equilibrium position of the shuttle) can
be stimulated by  Kondo tunneling.  The latter case demonstrates the
potential for a Kondo induced electromechanical instability.

\begin{figure}
\includegraphics[width=7cm]{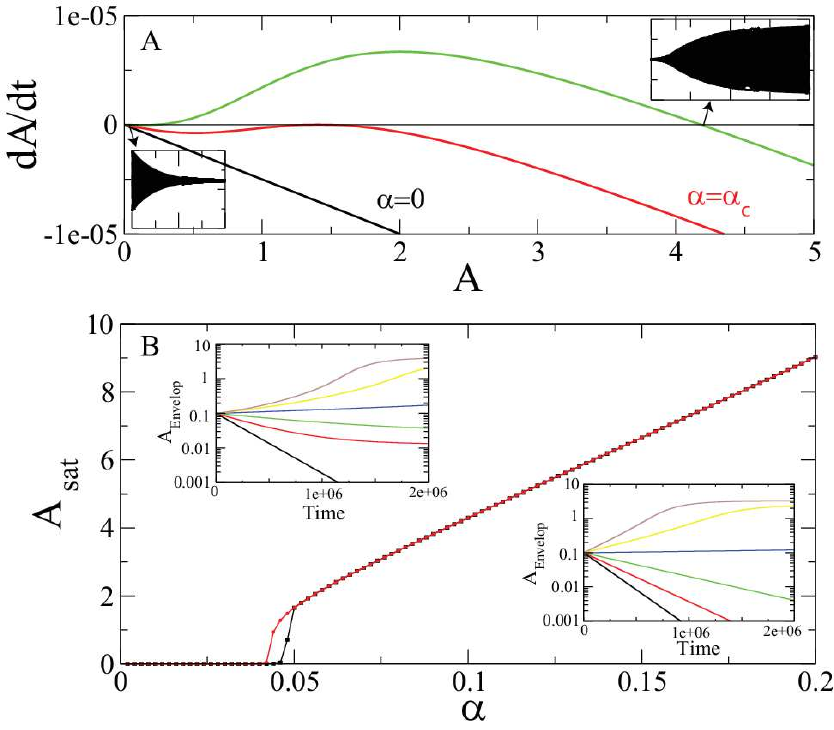}
\vspace{5mm}
 \caption{ Panel A: Amplitude dynamics at different values of the dimensionless force $\alpha$
  (see details in the text). Insets: time trace of the oscillation at two different fixed point indicated by arrow.
   { Panel B}: Saturation amplitude as a function of dimensionless force. Different colors denote initial conditions near (black dots) and far (red dots) from the equilibrium position $y_{0}$.
Insets: {amplitude envelope} as a function of dimensionless time
calculated by using Eq. (\ref{force}). The parameter $\alpha$ varies
from $\alpha=0$ (black) to $\alpha=0.1$ (magenta). The equations are
solved for the following set of parameters: $\beta=8$,
$\gamma=10^{-5}$, {$y_{0}=0.5$} and
$\frac{\hbar\omega_{0}}{k_{B}T_{K}^{min}}=10^{-3}$. Reprinted with permission from
\cite{kis12}, T. Song {\em et al.}, New Journal of Physics, {\bf 16},
033043 (2014).}
 \label{fig:amp}
\end{figure}

{ In order to describe these instability, one should discuss the
contribution of "Kondo force" $F_K$ to the right hand side part
(\ref{4.6}) of Eq. (\ref{2.4}). This force consists of two
components \cite{Korean}:
\begin{equation}
F_K = -\frac{\alpha_K +\alpha_{\rm
ret}}{\cosh^2(y-y_0)\omega_0^2\lambda}.
\end{equation}
where
\begin{eqnarray}\label{force}
\alpha_K &=&\frac{\pi E_Ck_BT_K(t)}{8\Gamma_0\lambda},\\
\alpha_{\rm ret}&=& 2\dot{y}G_0V_{\rm bias}BL\tanh(y-y_0)\tau_{\rm
ret}e^{-\beta[1+\tanh(y-y_0)]/2}\nonumber
\end{eqnarray}
Here $\beta=\pi E_C/4\Gamma_0$ is the coupling strength of
electronic states.
 The first term stems from the Kondo cloud adiabatically following the change of $T_K(t)$
induced by the moving shuttle in the source electrode and metallic
cantilever. The second term describes the temporal retardation
related to dynamics of Kondo cloud with the characteristic time
$\tau_{ret}= \hbar\omega_0\beta/(2k_BT_K^{\rm min})$. The time
dependent Kondo temperature in the strong coupling limit at $T\ll
T_{K}^{min}$ is given by
\begin{eqnarray}
k_{B}T_{k}(t) =
k_{B}T_{K}^{min}\exp\left\{\frac{\beta}{2}[1+\tanh(y(t)-y_{0})])\right\}.
\end{eqnarray}
The $k_{B}T_{K}^{min}$ plays the role of the cutoff energy for Kondo
problem.

 The instability is controlled by the bias $V_{\rm bias}$ entering $\alpha_{\rm ret}$.
 Fig. \ref{fig:amp} illustrates two regimes of Kondo shuttling. Namely, at
 small bias the Kondo force controlled by external fields further damps the oscillator,
 and we obtain an efficient mechanism of cooling the nano-shuttle. On the other hand,
 at $V_{\rm bias}$ above some treshold value, the contribution of the Kondo force
enhances the oscillations, and we arrive at the non-linear steady
state regime of self sustained oscillations. }

Summarizing,  we emphasize that the Kondo phenomenon in single
electron tunneling \color{black} gives a very promising and
efficient mechanism \color{black} for electromechanical transduction
on a nanometer length scale. Measuring the nanomechanical response
on Kondo-transport in a nanomechanical single-electron device
enables one to study the kinetics of the formation of
Kondo-screening and offers a new approach for studying
nonequilibrium Kondo phenomena.  The Kondo effect provides a
possibility for super high tunability of the mechanical dissipation
as well as super sensitive detection of mechanical displacement.

\section{Conclusions}

 {During the last several years there has been
 significant activity in the study of nanoelectromechanical (NEM) shuttle structures.
 In this review we concentrate on description of the influence of spin-related effects on the functionality of shuttle devices. In particular, we emphasize the importance of electronic spin in shuttle devices made of magnetic materials. }
 Spin-dependent exchange forces can be responsible for a
qualitatively new nanomechanical performance opening a new field of
study that can be called spintro-mechanics. Electronic many-body
effects, appearing beyond the weak tunneling approach, result in
single electron shuttling assisted by Kondo-resonance electronic
states. The possibility to achieve a high
sensitivity to coordinate displacement in electromechanical
transduction along with the possibility to study the kinetics of the
formation of many-body Kondo states has also been demonstrated.

There are still a number of unexplored shuttling regimes and
systems, which one could focus on in the nearest future. In addition
to magnetic shuttle devices one could explore hybrid structures
where the source/drain and gate electrodes are hybrids of magnetic
and superconducting materials. Then one could expect
spintromechanical actions of a supercurrent flow as well as
superconducting proximity effects in the spin dynamics in magnetic
NEM devices. An additional direction is the study of shuttle
operation under microwave radiation. In this respect microwave
assisted spintromechanics is of special interest due to the
possibility of microwave radiation to resonantly flip electronic
spins. As in ballistic point contacts such flips can be confined to
particular locations by the choice of microwave frequency, allowing
for external tuning of the spintromechanical dynamics of the
shuttle.

\section{Acknowledgements} \label{c}

Financial support from the Swedish VR, and the Korean WCU program
funded by MEST/NFR (R31-2008-000-10057-0) is gratefully
acknowledged. This research was supported in part by the Project of
Knowledge Innovation Program (PKIP) of Chinese Academy of Sciences,
Grant No. KJCX2.YW.W10. I. V. K. and A. V. P. acknowledge financial
support from the National Academy of Science of Ukraine (grant No.
4/13-N). I. V. K. thanks the Department of Physics at the University
of Gothenburg for hospitality.

\end{document}